\documentclass[aps,prb,twocolumn,floatfix,superscriptaddress,showpacs]{revtex4}
\usepackage{graphicx}
\usepackage{mathrsfs} 
\usepackage[mathscr]{euscript}
\usepackage{amsmath,amssymb}
\usepackage{bm}
\usepackage{color}
\usepackage{dcolumn}

\def\Tr{{\rm Tr}}

\def\sgn{{\rm sgn}}
\def\vec#1{ {\bm{#1}} }

\def\av_expe#1{ \langle #1 \rangle_{\rm av} }
\definecolor{purple}{rgb}{0.62745098,0.125490196,0.941176471}

\begin{document}
\title{Vector-spin-chirality order in a dimerized frustrated spin-$1/2$ chain}
\author{Hiroshi Ueda}
\affiliation{Condensed Matter Theory Laboratory, RIKEN, Wako, Saitama 351-0198, Japan}
\author{Shigeki Onoda}
\affiliation{Condensed Matter Theory Laboratory, RIKEN, Wako, Saitama 351-0198, Japan}
\affiliation{Quantum Matter Theory Research Team, CEMS, RIKEN, Wako, Saitama 351-0198, Japan}

\date{\today}

\begin{abstract}
A frustrated spin-$1/2$ XXZ chain model comprising a ferromagnetic nearest-neighbor coupling with the bond alternation, $J_1(1\pm\delta)<0$, and an antiferromagnetic second-neighbor exchange coupling $J_2>0$ is studied at zero and weak magnetic fields by means of density matrix renormalization group calculations of order parameters, correlation functions and the entanglement entropy as well as an Abelian bosonization analysis. At zero magnetic field, the bond alternation $\delta>0$ suppresses the gapless phase characterized by a vector-chiral (VC) long-range order (LRO) and a quasi-LRO of an incommensurate spin spiral, whereas this phase occupies a large region in the space of $J_1/J_2$ and the easy-plane exchange anisotropy for $\delta=0$  [S. Furukawa \textit{et al.}, Phys. Rev. Lett. \textbf{105}, 257205 (2010)]. Then, four gapped phases are found to appear as the exchange anisotropy varies from the SU(2)-symmetric case to the U(1)-symmetric case; the Haldane dimer (D$_+$) phase with the same sign of the $x,y$- and $z$-component dimer order parameters, two VC dimer (VCD$_+$/VCD$_-$) phases with the sign of the $z$-component dimer order parameter being unaltered/reversed, and the even-parity dimer (D$_-$) phase. At small magnetic fields, a field-induced ring-exchange interaction, which is proportional to a staggered scalar chirality and a magnetic flux penetrating the associated triangle, drives a transition from the D$_-$ phase into a VC-Neel-dimer (VCND) phase, but not from the D$_+$ phase. This VCND phase is stable up to the large magnetic field at which the Zeeman term closes the spin gap. A possible relevance to Rb$_2$Cu$_2$Mo$_3$O$_{12}$ is discussed.
\end{abstract}


\maketitle
\section{Introduction}
\label{sec:intro}

A novel spontaneously symmetry-broken state characterized by a long-range order (LRO) of a vector chirality $\sum_i\langle \hat{\bm{S}}_i \times \hat{ \bm{S} }_{i+1}\rangle\ne0$ of nearby spins $\hat{\bm{S}}_i$ in the absence of any magnetic spiral LRO~\cite{Villain77,Miyashita84,Kawamura98} has long been sought both theoretically and experimentally in frustrated magnets~\cite{Lecheminant05}. This order breaks the inversion symmetry but preserves the time-reversal, and can be detected as the linearly coupled ferroelectric polarization in insulating magnets~\cite{Katsura05,Sergienko06,Mostovoy06,Harris06,Jia08}. It has recently been shown that the order appears in a frustrated spin-$1/2$ chain with an antiferromagnetic second-neighbor exchange coupling ($J_2>0$) and weak easy-plane anisotropy~\cite{Nersesyan98,Lecheminant01,Hikihara01,furukawa08,FSO10}, particularly on the ferromagnetic side of the nearest-neighbor exchange coupling ($J_1<0$)~\cite{furukawa08,FSO10}. However, this state is fragile in three dimensions, since it is mostly accompanied by a quasi-LRO of spins with gapless excitations~\cite{Nersesyan98} and is eventually driven to a magnetic spiral LRO by an interchain coupling. Indeed, such spin-spiral LRO and the associated ferroelectric polarization have been observed in relevant cuprate compounds, LiCuVO$_4$~\cite{Enderle05,Naito07}, LiCu$_2$O$_2$~\cite{Masuda05,Park07}, and Pb[Cu(SO$_4$)(OH)$_2$]~\cite{Yasui11,Wolter012}. Therefore, a spin excitation gap is required for protecting the vector-spin-chirality ordered state without a magnetic spiral LRO.

With similar values of $J_1/J_2$ ($\sim-2.7$) and weak three-dimensional couplings, however, Rb$_2$Cu$_2$Mo$_3$O$_{12}$ provides another prototype of spin-$1/2$ chain. It does not show a spontaneous symmetry breaking down to 2~K~\cite{Hase04}. Each spin chain in this compound actually has an alternation in $J_1$ due to a weak crystallographic dimerization~\cite{Solodovnikov97}. It is natural that the zero-field ground state will be adiabatically connected to either of the following two dimer phases obtained in Ref.~\onlinecite{FSOF12}, which we will classify as distinct symmetry-protected topological phases~\cite{Pollmann10, Liu12} in the absence of the Neel order. (i) One is a Haldane dimer state~\cite{FSOF12}, that is, a Haldane state~\cite{Affleck87,Haldane83} formed by spin triplet pairs of the ferromagnetically-coupled nearest-neighbor spins, as it appears in the vicinity of the SU(2)-symmetric case~\cite{Itoi01,FSOF12}. (ii) The other is an even-parity dimer state realized with large enough easy-plane magnetic anisotropy.  Furthermore, recent experiments have shown that the applied magnetic field in a range of 0.1--2.0~T induces a ferroelectric polarization in the absence of a magnetic spiral LRO without closing the spin gap~\cite{Yasui13}, possibly realizing the genuine spin-gapped vector-spin-chirality ordered state. Such vector-chiral ordered state can be realized under a weak Zeeman field~\cite{Kolezhuk05,Hikihara08}. In this case, however, it is accompanied by a quasi-LRO of spins~\cite{Kolezhuk05,Hikihara08} which should evolve into a magnetic spiral LRO in three dimensions, in contrast to the experimental results~\cite{Yasui13}. Therefore, a theoretical scenario for such (field-induced) gapped vector-chiral (VC) state is called for. 

Here, we investigate a frustrated spin-$1/2$ XXZ chain with a bond alternation using the density matrix renormalization group (DMRG) method~\cite{White:PRL69_White:PRB48, Peschel:Springer, doi:10.1142/S0217979299000023, RevModPhys.77.259, Schollwock201196} and a bosonization analysis. It is shown that a small bond alternation does not completely kill the vector-spin-chirality order but drives the otherwise gapless VC phase into two gapped VC phases. A finite Zeeman field induces a transition to a gapless VC phase but not to a gapped VC phase. We propose a vital role of a field-induced ring-exchange interaction taking the form of a staggered scalar chirality. It is found that it generates a gapped VC phase accompanied by a tiny Neel order, intervening between a dimer phase and gapless VC phase with increasing magnetic field. 

We start with the case of zero magnetic field in Sec.~\ref{sec:0mag}. Then, we reveal a gapped VC dimer state induced by a staggered scalar chiral interaction under a weak field in Sec.~\ref{sec:ssc}. An Abelian bosonization analysis is performed in Sec.~\ref{sec:bosonization} to understand gapped VC phases, namely, two VC dimer phases and a field-induced VC Neel dimer phase, as well as transitions among them. We then give in Sec.~\ref{sec:Zeeman} a phase diagram that involves a gapless VC state induced by the Zeeman field  with further increasing magnetic field. Section~\ref{sec:conclustion} is devoted to the conclusion and discussions on a relevance to Rb$_2$Cu$_2$O$_3$Mo$_{12}$.
The readership interested in only the conclusions on the phase diagrams but not details of the numerics can skip to Fig.~\ref{fig:V=h=0} for the zero-field case, Fig.~\ref{fig_delta_vh} for the low-field case, and Fig.~\ref{fig:h_1ov_D708} for the high-field case.

\section{Zero field case} 
\label{sec:0mag}

We start from a spin-$1/2$ XXZ model with the bond alternation in the ferromagnetic nearest-neighbor exchange coupling $J_1<0$ at zero magnetic field, as schematically shown in Fig.~\ref{fig:V=h=0}~(a),
\begin{eqnarray}
\hat{ \mathcal{H} }_{\delta\mathrm{xxz}}
 &=& J_1 \sum_{j=1}^{N-1} ( 1 - (-1)^j \delta )\left[\sum_{\mu=x,y}\hat{S}^\mu_{j} \hat{S}^\mu_{j+1} + \Delta \hat{S}^{z}_{j} \hat{S}^{z}_{j+1}\right] \nonumber \\
&&+ J_2 \sum_{j=1}^{N-2} \left[\sum_{\mu=x,y}\hat{S}^\mu_{j} \hat{S}^\mu_{j+2} + \Delta \hat{S}^{z}_{j} \hat{S}^{z}_{j+2}\right],
\label{eq:Hdxxz}
\end{eqnarray}
where $\hat{S}^\mu_j$ with $\mu=x,y,z$ denotes the spin-$1/2$ operator at the site $j$, and $\delta$ and $\Delta$ represent the relative amplitude of the bond alternation and the easy-plane anisotropy ($0\le\Delta\le1$), respectively.  We take the number $N$ of spins to be even and $\delta\ge0$ so that the nearest-neighbor coupling terminates with the stronger amplitude $J_1(1+\delta)$ at the both edges.

\subsection{Phase diagrams}
\label{sec:0mag:phase diagram}
\begin{figure}[H!tb]
  \centering
  \resizebox{7.5cm}{!}{\includegraphics{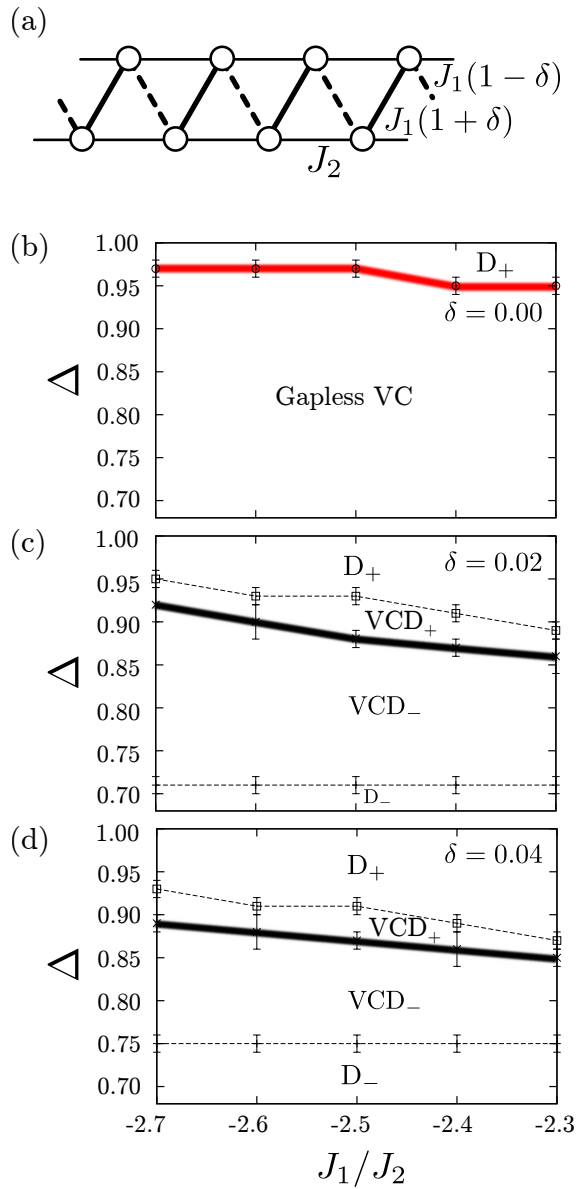}}
  \caption{ 
(Color online) The lattice structure (a) and the phase diagrams of $\hat{ \mathcal{H} }_{\delta\mathrm{xxz}}$ (Eq.~\eqref{eq:Hdxxz}) for (b) $\delta = 0$, (c) $\delta = 0.02$, and (d) $\delta = 0.04$. The phase diagrams are obtained with DMRG calculations up to the system size $N=320$ and the 300 renormalized basis states ($m=300$). We have also checked a convergence of the results by increasing $m$ up to 500 in the case of $J_1/J_2=-2.5$ and $\delta=0.02$. A possible VC Haldane dimer (VCD$_+$) phase is restricted inside the red shaded area at most in (b). A possible gapless VC dimer (VCD$_0$) phase with only the $x,y$-component dimer order is restricted inside and the black shaded area at most in (c) and (d).
}
\label{fig:V=h=0}
\end{figure}

First, we summarize the zero-field ground-state phase diagrams for $\delta=0$, 0.02, and 0.04 in Fig.~\ref{fig:V=h=0}~(b), (c), and (d), respectively, which are obtained by carefully examining correlation functions (Fig.~\ref{fig:col_V=h=0}) and entanglement entropy (Fig.~\ref{fig:EE}), which we will discuss later. Without the bond alternation, i.e., $\delta=0$, our DMRG calculations mostly reproduce the previously obtained phase diagram~\cite{FSOF12} (Fig.~\ref{fig:V=h=0}~(b)). Around the U(1)-symmetric case $\Delta=0$, there appears the even-parity dimer (D$_-$) phase, characterized by the opposite signs of the $z$-component and $x$- or $y$-component (inplane) dimer ordering amplitudes, i.e., $\langle \hat{D}^x_j\rangle\langle \hat{D}^z_j\rangle<0$, for every $j$ away from the edges, where
\begin{equation}
\hat{D}^{\mu}_{j} = (-1)^j(\hat{S}^{\mu}_{j-1} \hat{S}^{\mu}_{j} - \hat{S}^{\mu}_{j} \hat{S}^{\mu}_{j+1}),
\end{equation}
is the $\mu$-component dimer operator.
With increasing $\Delta$, the gapless VC phase with a finite $z$-component of the VC order parameter, $\mathrm{Ave}_j\langle\hat{\kappa}^z_{j+1/2}\rangle\ne0$, appears in a wide range in the space of $(J_1/J_2<0,0\le\Delta\le1)$, where
\begin{equation}
\hat{\kappa}^{z}_{j+1/2} = ( \hat{\vec{S}}_j \times \hat{ \vec{S} }_{j+1} )^{z}, 
\end{equation}
and $\mathrm{Ave}_j$ denotes a spatial average only over $N/2$ spins located at the center of the whole chain, which is required for excluding the effects of the boundaries. (See Appendix~\ref{average_process} for details on how it works.)
The gapless VC phase is eventually taken place by the Haldane dimer (D$_+$) phase as the SU(2)-symmetric line ($\Delta=1$) is approached. In this D$_+$ phase, two dimer ordering amplitudes $\mathrm{Ave}_j\langle\hat{D}^{x,y}_j\rangle$ and $\mathrm{Ave}_j\langle\hat{D}^z_j\rangle$ have the same sign. There are also narrow intermediate phases, i.e., a VCD$_+$ phase where the VC order coexists with the Haldane dimer order and a VCD$_-$ phase where the VC order coexists with the even-parity dimer order. Actually, it has been shown in Ref.~\onlinecite{FSOF12} that the VCD$_+$ phase appears only inside the red shaded area in Fig.~\ref{fig:V=h=0}~(b), though its detection is beyond the current numerical accuracy. 

A small bond alternation $\delta\ne0$ narrows the VC ordered region, and endows an energy gap in spin excitations in an otherwise gapless VC phase, yielding the two VC dimer (VCD$_+$ and VCD$_-$) phases, as shown in Fig.~\ref{fig:V=h=0}~(c) and (d) for $\delta=0.02$ and $0.04$, respectively. It is noteworthy that two distinct VC dimer phases, denoted as VCD$_+$ and VCD$_-$, are separated by a boundary at which the energy gap is closed. This gapless phase boundary either occurs at a line or form a stable gapless VC dimer phase with only the $x$, $y$-component dimer ordering amplitudes in a very narrow region as shown with black shades in Fig.~\ref{fig:V=h=0}~(c) and (d).

\subsection{Correlation functions}
\label{sec:0mag:correlation}
\begin{figure*}[H!tb]
  \centering
  \resizebox{17.6cm}{!}{\includegraphics{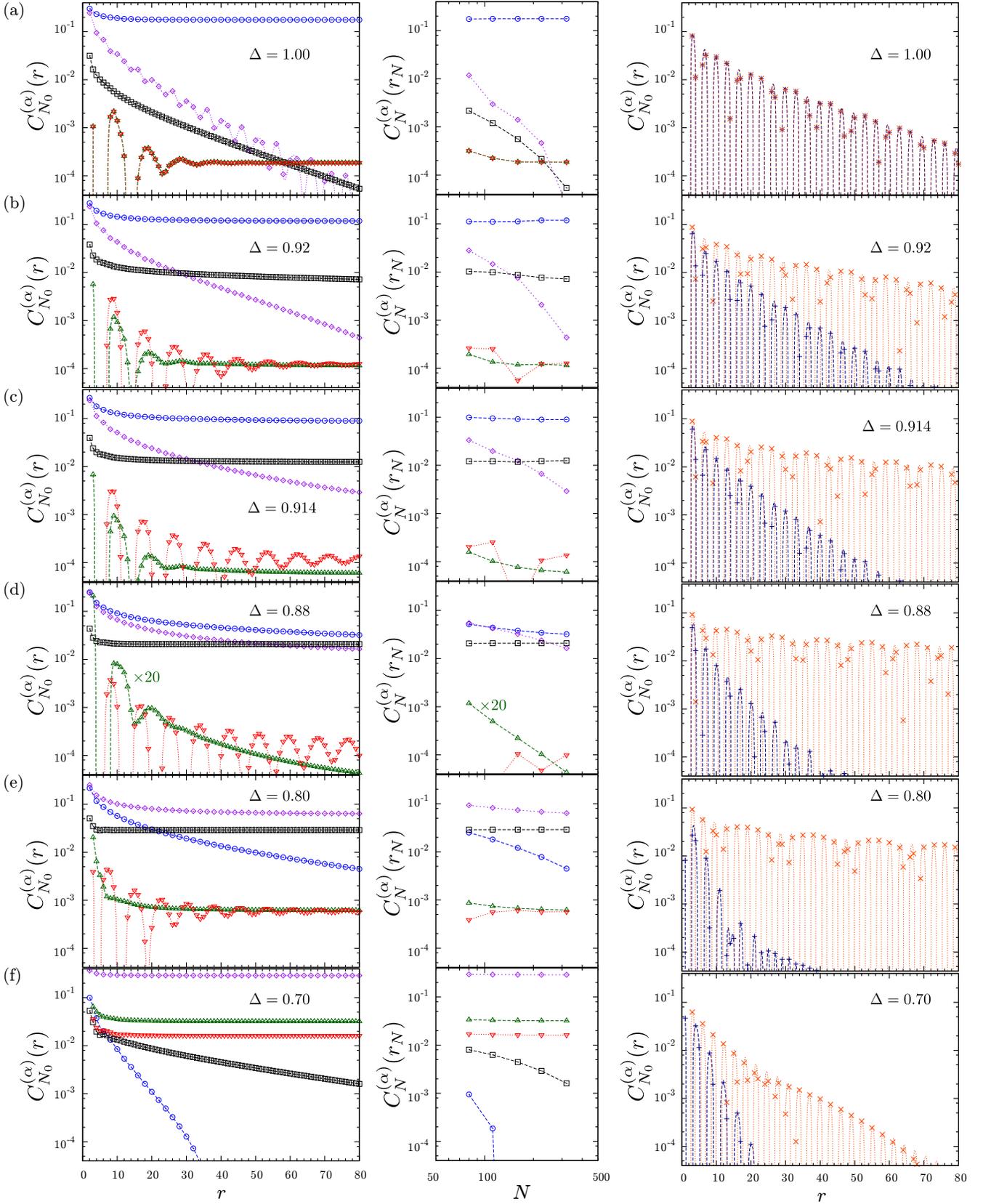}}
  \caption{ 
(Color online) Correlation functions $C^{(\alpha)}_{N}(r)$ for (a) $\Delta=1.00$ in the Haldane dimer (D$_+$) phase, (b) $\Delta=0.92$ near the boundary between D$_+$ phase and VCD$_+$ phase, (c) $\Delta=0.914$ in the VCD$_+$ phase, (d) $\Delta=0.88$ near the boundary between VCD$_+$ phase and VCD$_-$ phase or in a very narrow gapless VCD$_0$ phase, (e) $\Delta=0.80$ in the VCD$_-$ phase, and (f) $\Delta=0.70$ in the even-parity dimer (D$_-$) phase in the case of $J_1/J_2=-2.5$ and $\delta=0.02$. The left and middle subpanels of (a)-(e) show the spatial decays of $C^{(\alpha)}_{N_0}(r)$ with $N_0=320$ and the $N$ dependence of $C^{(\alpha)}_{N}(r_N)$ with $r_N = N/4$ for $N=80, 112, 160, 224$ and $320$, respectively ($C^{(O^z_1)}_{N}(r):\textcolor{blue}{\bigcirc}$, $C^{(O^z_2)}_{N}(r):\textcolor{purple}{\Diamond}$, $C^{(\kappa^z)}_{N}(r):\Box$, $C^{(D^z)}_{N}(r):\textcolor{green}{\bigtriangleup}$, $C^{(D^x)}_{N}(r):\textcolor{red}{\bigtriangledown}$). The right subpanels show the spatial decay of spin correlation functions ($C^{(S^z)}_{N}(r):\textcolor{blue}{+}$, $C^{(S^x)}_{N}(r):\textcolor{red}{\times}$). The results are obtained with $m=500$, and are almost unchanged within the symbol size by taking $m=300$.
For a guide to the eyes, the cubic-spline interpolated curve is plotted for each correlation function. 
}
\label{fig:col_V=h=0}
\end{figure*}

To identify the phases shown in Fig.~\ref{fig:V=h=0}~(b)-(d), we have examined string, VC, dimer, and (staggered) spin correlation functions,
\begin{align}
C^{(O^z_\ell)}_{N}(2r) &= \mathrm{Ave}_j \langle \hat{O}^{z}_{2j+\ell,2j+\ell+2r} \rangle
~~~(\ell=1,2),
\notag\\
C^{(\kappa^{z})}_{N}(r) &= \mathrm{Ave}_j \langle \hat{\kappa}^{z}_{j+1/2} \hat{\kappa}^{z}_{j+1/2+r} \rangle,
\notag\\
C^{(D^{\mu})}_{N}(r) &= \mathrm{Ave}_j \langle \hat{D}^{\mu}_{j} \hat{D}^{\mu}_{j+r} \rangle
~~~(\mu=x,y,z),
\notag\\
C^{(S^{\mu})}_{N}(r) &= (-1)^r \mathrm{Ave}_j \langle \hat{S}^{\mu}_{j} \hat{S}^{\mu}_{j+r} \rangle
~~~(\mu=x,y,z),
\label{eq:correlaton_functions}
\end{align}
with the nonlocal string operator~\cite{denNijs89,Tasaki91,Watanabe93,Nishiyama95,White96_ladder,Kim00}
\begin{align}
\hat{O}^{z}_{j,j+2r} &= - (\hat{S}^{z}_{j} + \hat{S}^{z}_{j+1}) \exp( i \pi \sum^{2r-1}_{k=2} \hat{S}^{z}_{j+k} )\nonumber \\ 
&\times (\hat{S}^{z}_{j+2r} + \hat{S}^{z}_{j+2r+1}).
\end{align}
Note that two string order parameters $\lim_{r\to\infty}C^{(O^z_\ell)}_{N}(2r)$  with $\ell=1,2$ give useful probes of two distinct dimer structures; 
\begin{align}
  \lim_{r\to\infty}C^{(O^z_2)}_{N}(2r)>0,
  ~
  \lim_{r\to\infty}C^{(O^z_1)}_{N}(2r)=0
  ~~\mbox{(D$_-$, VCD$_-$)}, 
  \notag\\
  \lim_{r\to\infty}C^{(O^z_2)}_{N}(2r)=0,
  ~
  \lim_{r\to\infty}C^{(O^z_1)}_{N}(2r)>0
  ~~\mbox{(D$_+$, VCD$_+$)}.
  \label{eq:string}
\end{align}

We have extensively performed DMRG calculations and computed the correlation functions, while increasing system size $N$ up to $N_0=320$. For each $N$, we have checked the convergence of the results with increasing number $m$ of renormalized basis states. Throughout this section, we show results obtained with $m = 500$. The results in the case of $J_1/J_2=-2.5$ and $\delta=0.02$ are shown for $N\le N_0$ in Fig.~\ref{fig:col_V=h=0}: the spatial ($r$) dependences of string, dimer, and VC correlation functions for $N=N_0$ are in the left panels, the $N$ dependence of these correlation functions at a long distance $r_N=N/4$ are in the middle panels, and the $r$ dependence of spin correlation functions for $N=N_0$ are in the right panels.

Figure~\ref{fig:col_V=h=0}~(a)  shows the results for $\Delta=1$, which is located in the Haldane dimer (D$_+$) phase~\cite{Itoi01,FSOF12}. A string correlation $C_N^{(O^z_1)}(2r)$ and dimer correlations $C_N^{(D^x)}(2r)$ and $C_N^{(D^z)}(2r)$ are saturated to finite values up to $r\sim80$ and $N=N_0=320$, while the other string correlation $C_N^{(O^z_2)}(2r)$, the VC correlation $C_N^{(\kappa^z)}(r)$, and the spin correlations $C_N^{(S^x)}(r)$ and $C_N^{(S^z)}(r)$ decay exponentially to zero with increasing $r$ (left and right panels). The saturation of a string correlation $C_N^{(O^z_2)}(2r)$ and dimer correlations $C_N^{(D^x)}(2r)$ and $C_N^{(D^z)}(2r)$ and the exponential decay of the VC correlation $C_N^{(\kappa^z)}(r)$ can also be confirmed by their behaviors at $r=N/4$ with increasing $N$ (the middle panels). We have also checked $\langle \hat{D}^{z}_j \rangle \langle \hat{D}^{x}_j \rangle > 0$ holds for $j$ away from the boundary region (Fig.~\ref{fig:scaling} (a)), as is strictly required by the SU(2) symmetry. This relation holds even away from the SU(2)-symmetric case $\Delta=1$. All the above properties are exclusively consistent with the Haldane dimer (D$_+$) phase. 

One can see a growth of the VC correlation with decreasing $\Delta$ from 1 to 0.92, approaching a phase transition to the VCD$_+$ phase. 
The VC correlation is indeed almost critical for $\Delta=0.92$, while the other properties do not change qualitatively except that the SU(2) symmetry is lost, as shown in Fig.~\ref{fig:col_V=h=0}~(b): it almost shows a power-law decay as is clear from both the left and the middle panels. In fact, the string correlation $C_N^{(O^z_1)}(2r)$ and dimer correlations $C_N^{(D^x)}(2r)$ and $C_N^{(D^z)}(2r)$ remain long-range for $\Delta=0.914$, as shown in Fig.~\ref{fig:col_V=h=0}~(c). The nature of the dimer order is the same as in the $D_+$ phase:  $\langle \hat{D}^{z}_j \rangle \langle \hat{D}^{x}_j \rangle > 0$ holds for $j$ away from the boundary regions. (See Fig.~\ref{fig:scaling} (a)). These findings indicate that the $\Delta=0.914$ case is located inside the VCD$_+$ phase. Since the $z$-component dimer correlation at a distance $r_{N_0}=80$ for $N=N_0=320$ decreases by a factor of 3.1 with decreasing $\Delta$ from 1 to 0.914, the energy gap in the inplane spin excitation is reduced. Accordingly, the inplane spin correlation length becomes longer than or comparable to $r_{N_0}=80$, as we see in the right panel.

Further decreasing $\Delta$, we find a minimum in the magnitude of the $z$-component dimer correlation around $\Delta=0.88$, as shown in Fig.~\ref{fig:col_V=h=0}~(d). In this case, the two string correlation functions and the $z$-component dimer correlation function all almost show a power-law decay, while the VC correlation and the $x$-component dimer correlation remain long-range (left and middle panels). The inplane spin correlation is also consistent with the almost gapless feature (right panel). 

This gapless feature around $\Delta=0.88$ actually signals a phase transition from the VCD$_+$ phase to a VCD$_-$ phase. This transition may occur either directly or in a narrow but finite region of a gapless VC dimer (VCD$_0$) phase with only $x,y$-component dimer correlations being long-range. Figure~\ref{fig:col_V=h=0}~(e) indicates that for $\Delta=0.80$, a long-range nature of the longitudinal dimer correlation is recovered but with a reversal of the relative sign of the dimer order parameters, so that $\langle \hat{D}^{z}_j \rangle \langle \hat{D}^{x}_j \rangle < 0$. Accordingly, the string correlation $C_N^{(O^z_2)}(r)$ becomes long-range with a larger order parameter amplitude than in the case of $\Delta=1$, while the other one $C_N^{(O^z_1)}(r)$ becomes short-range, showing an exponential decay. The VC correlation remains long-range. (See the left and middle panels.) However, the inplane spin correlation still has a long correlation length (right panel).

Further decreasing $\Delta$ to 0.70, the system is now in the even-parity dimer (D$_-$) phase. As shown in Fig.~\ref{fig:col_V=h=0}~(f), the VC correlation eventually shows an exponential decay and becomes short-range, while the long-range nature of the dimer correlations remain the same as in the case of $\Delta=0.80$, but with even larger order parameter amplitudes (left and middle panels). Accordingly, the spin correlations are clearly all short-range (right panel).

\subsection{Order parameters}
\label{sec:0mag:entanglement}
\begin{figure}[H!tb]
  \centering
  \resizebox{8.6cm}{!}{\includegraphics{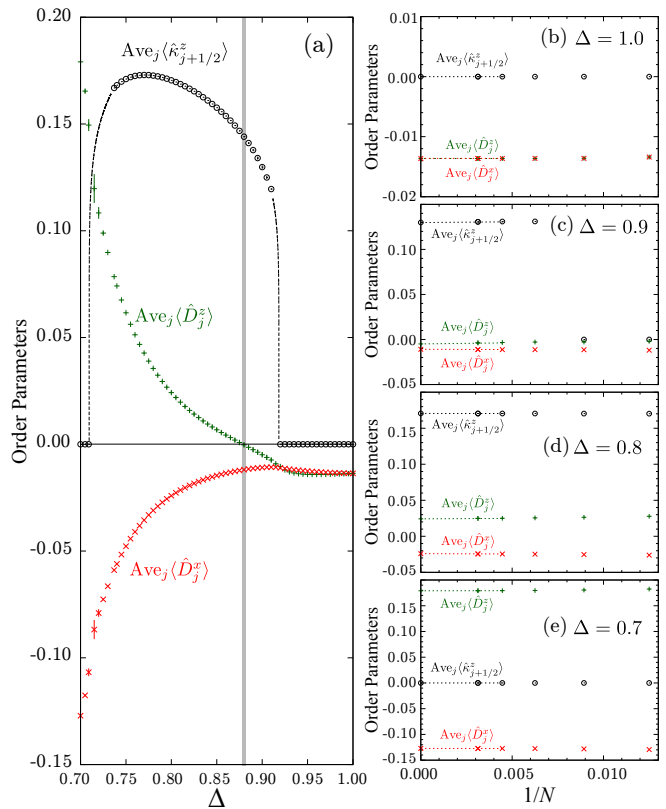}}
  \caption{ (Color online) 
(a) Bulk order parameters $\mathrm{Ave}_j\langle\hat{\kappa}^z_{j+1/2}\rangle$, $\mathrm{Ave}_j\langle\hat{D}^x_j\rangle$, and $\mathrm{Ave}_j\langle\hat{D}^z_j\rangle$ extrapolated to $N \rightarrow \infty$ for $J_1/J_2=-2.5$ and $\delta=0.02$. See the text for the broken lines and the gray line. (b-e) Order parameters for $N=80$, $112$, $160$, $224$, and $320$. The order parameters in the bulk limit are estimated by a linear extrapolation of the data for $N=224$ and $320$. The error bars in the panel (a) are chosen to be the difference between the data for $N=320$ and the extrapolated value. All the results are obtained with $m=500$ and are unchanged by taking $m=400$ within the symbol size.
}
\label{fig:scaling}
\end{figure}

The emergence of the VC LRO is also indicated by an observation of the order parameter $\mathrm{Ave}_j\langle\hat{\kappa}^z_{j+1/2}\rangle$ with the technique developed by Okunishi~\cite{Okunishi08}. Order parameters for $J_1/J_2=-2.5$ and $\delta=0.02$ in the thermodynamic limit are shown in Fig.~\ref{fig:scaling} (a), where all the plotted symbols are obtained by linearly extrapolating the data for $N = 224$ and 320 to $1/N \rightarrow 0$ as exemplified in Fig.~\ref{fig:scaling} (b-e). The VC order parameter steeply decays near the D$_-$-VCD$_-$ phase boundary ($\Delta_{ {\rm D}_- {\rm \mathchar`-VCD}_-}$) and the D$_+$-VCD$_+$ phase boundary ($\Delta_{ {\rm D}_+ {\rm \mathchar`-VCD}_+}$). From a careful analysis given in Appendix \ref{fss_vc}, it has been found that the critical points are located at $\Delta_{ {\rm D}_- {\rm \mathchar`-VCD}_-} \simeq 0.709$ and $\Delta_{ {\rm D}_+ {\rm \mathchar`-VCD}_+} \simeq 0.918$. Namely, we can conclude $\mathrm{Ave}_j\langle\hat{\kappa}^z_{j+1/2}\rangle$ is finite for $0.709<\Delta<0.918$, indicating a VC LRO in this region. The sign change in the $z$-component dimer order parameter relative to the $x$-component can also be clearly seen at $\Delta\simeq0.88$: $\mathrm{Ave}_j\langle\hat{D}^x_j\rangle\mathrm{Ave}_j\langle\hat{D}^z_j\rangle$ is positive for $\Delta \gtrsim 0.88$ while it is negative for $\Delta \lesssim 0.88$. These observations are fully consistent with the behavior of the correlation functions (Fig.~\ref{fig:col_V=h=0}) and the phase diagram (Fig.~\ref{fig:V=h=0}~(c)), which we have already explained in Sec.~\ref{sec:0mag:correlation}.

It might be possible that the $z$-component dimer order parameter $\mathrm{Ave}_j\langle\hat{D}^z_j\rangle$ vanishes in an extended region, yielding a finite gapless VCD$_0$ phase. 
From the accuracy of the extrapolation, we can safely conclude that the gapless VCD$_0$ state does not extend outside the region of $0.879 \leq \Delta \leq  0.883$ (shown with a gray line in Fig.~\ref{fig:scaling} (a)) or just appears at a single point. This region for a gapless VCD$_0$ state is also consistent with the results found from the behavior of z-component of dimer correlations discussed in Sec.~\ref{sec:0mag:correlation}.

\subsection{Central charge}
\label{sec:0criticality}
\begin{figure}[]
  \centering
  \resizebox{8.5cm}{!}{\includegraphics{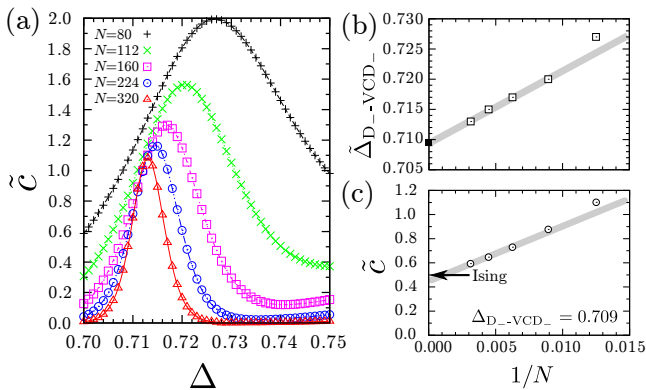}}
  \caption{ (Color online)
(a) Finite size dependence of $\tilde{c}$ around the D$_-$-VCD$_-$ boundary ($\Delta_{ {\rm D}_- {\rm \mathchar`-VCD}_-}$) for $J_1/J_2=-2.5$ and $\delta = 0.02$ ($m=500$). 
(b) The peak position $\tilde{\Delta}_{ {\rm D}_- {\rm \mathchar`-VCD}_-}$ of $\tilde{c}$ versus $1/N$ are open squares. The filled symbol represents $\Delta_{ {\rm D}_- {\rm \mathchar`-VCD}_-}$ determined from  Fig.~\ref{fig:scaling} (a). (c) Finite size dependence of $\tilde{c}$ at the criticality $\Delta_{ {\rm D}_- {\rm \mathchar`-VCD}_-}$.  The gray lines are a guide to the eyes. 
}
\label{fig:EE}
\end{figure}
A criticality in the phase diagram can also be probed through the von Neumann entanglement entropy of the right or left half of the spin chain, $S^{\mathrm{vN}} = -\Tr_A[\rho_{A} \log \rho_{A}]$ with $\rho_{A}$ being a reduced density matrix of either half. It scales with $\frac{c}{6}\log\frac{\xi}{\pi} +S_0$ around a criticality or with $\frac{c}{6}\log \frac{N}{\pi} +S_0$ at the criticality for $N\to\infty$, where $c$ is a central charge of the associated conformal field theory, $\xi$ is the relevant correlation length, and $S_0$ is a non-universal constant~\cite{Vidal03,Calabrese04}. 
Actually, the entanglement entropy depends on the position at which the system is split into two. In particular, at criticality, it behaves as 
\begin{equation}
S_A(N, \ell) = \frac{\tilde{c}}{6} \log\left[\frac{N}{\pi}\sin\frac{\pi \ell}{N}\right] + S_0,
\label{eq:S_A}
\end{equation}
in the open boundary condition~\cite{Calabrese04}, where $\ell$ is the number of spins involved in either side of two split subsystems. The coefficient $\tilde{c}$ asymptotically approaches the central charge $c$ for the criticality as $N\to\infty$. Note that $S^{\mathrm{vN}}$ is equal to $S_A$ of $\ell = N/2$. 
On the other hand, in an off-critical region where the longest length scale or correlation length $\xi$ of the system is finite, the entropy $S_A$ becomes saturated and thus $\tilde{c}$ should decay to zero as $\ell$ and $N$ increase much larger than $\xi$. First, we fit the numerical results of $S_A(N, \ell)$ at $\ell = N/2 - 18, N/2 - 16, \cdots, N/2$ to Eq.~\eqref{eq:S_A} with two adjustable parameters $\tilde{c}$ and $S_0$ for each $N$. Note that the spin chain is split into two at a weak bond in this case. The fitting procedure is found to be successful around the D$_-$-VCD$_-$ boundary. The result obtained for $\tilde{c}$ is shown in Fig.~\ref{fig:EE} (a).
Pursuing the peak position in $\tilde{c}$ up to $N=320$, it is clear from Fig.~\ref{fig:EE}~(b) that it is extrapolated to the almost same critical value $\Delta_{{\rm D}_- {\rm \mathchar`-VCD}_-} \simeq 0.709$ as we obtained by analyzing the VC order parameter in Sec. \ref{sec:0mag:entanglement}. (See also Appendix \ref{fss_vc}.)
 
Then, we can see that the value of $\tilde{c}$ for the critical $\Delta$ reasonably converges to the value 1/2 of the Ising universality class from Fig.~\ref{fig:EE}~(c).
These analyses are not successful around $0.85 \leq \Delta \leq 0.95$, which includes VCD$_-$-VCD$_0$, VCD$_0$-VCD$_+$, and D$_+$-VCD$_+$ boundaries. It is because the system size is too small to discuss the critical properties in this region.

\section{Low-field case without magnetization: staggered scalar chiral interaction}
\label{sec:ssc}
\begin{figure}[H!tb]
  \resizebox{7.5cm}{!}{\includegraphics{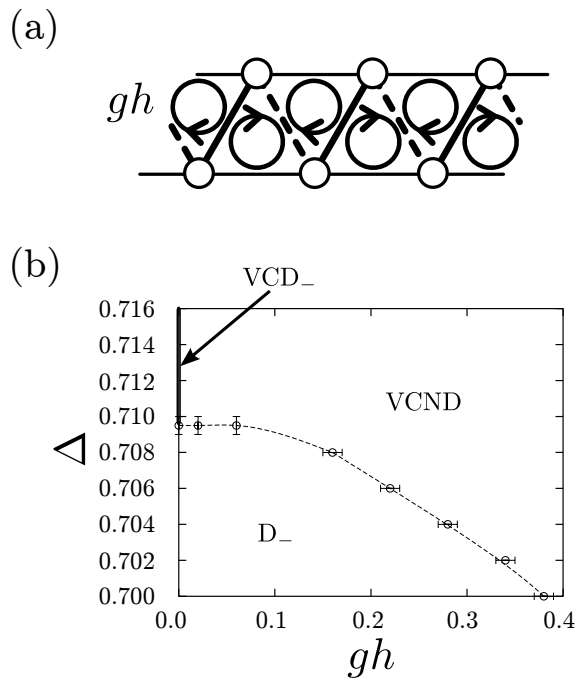}}
  \caption{ 
(a) Schematic picture of the staggered scalar spin chirality and (b) phase diagram of $\hat{ \mathcal{H} }_{\delta\mathrm{xxz}} + \hat{ \mathcal{H} }_{\mathrm{ssc}}$ around $\Delta = 0.7$ for $J_1/J_2 = -2.5$ and $\delta=0.02$.
The number $m$ of renormalized basis state in the DMRG method is 300.
}
\label{fig_delta_vh}
\end{figure}
\begin{figure*}[H!tb]
  \centering
  \resizebox{17.8cm}{!}{\includegraphics{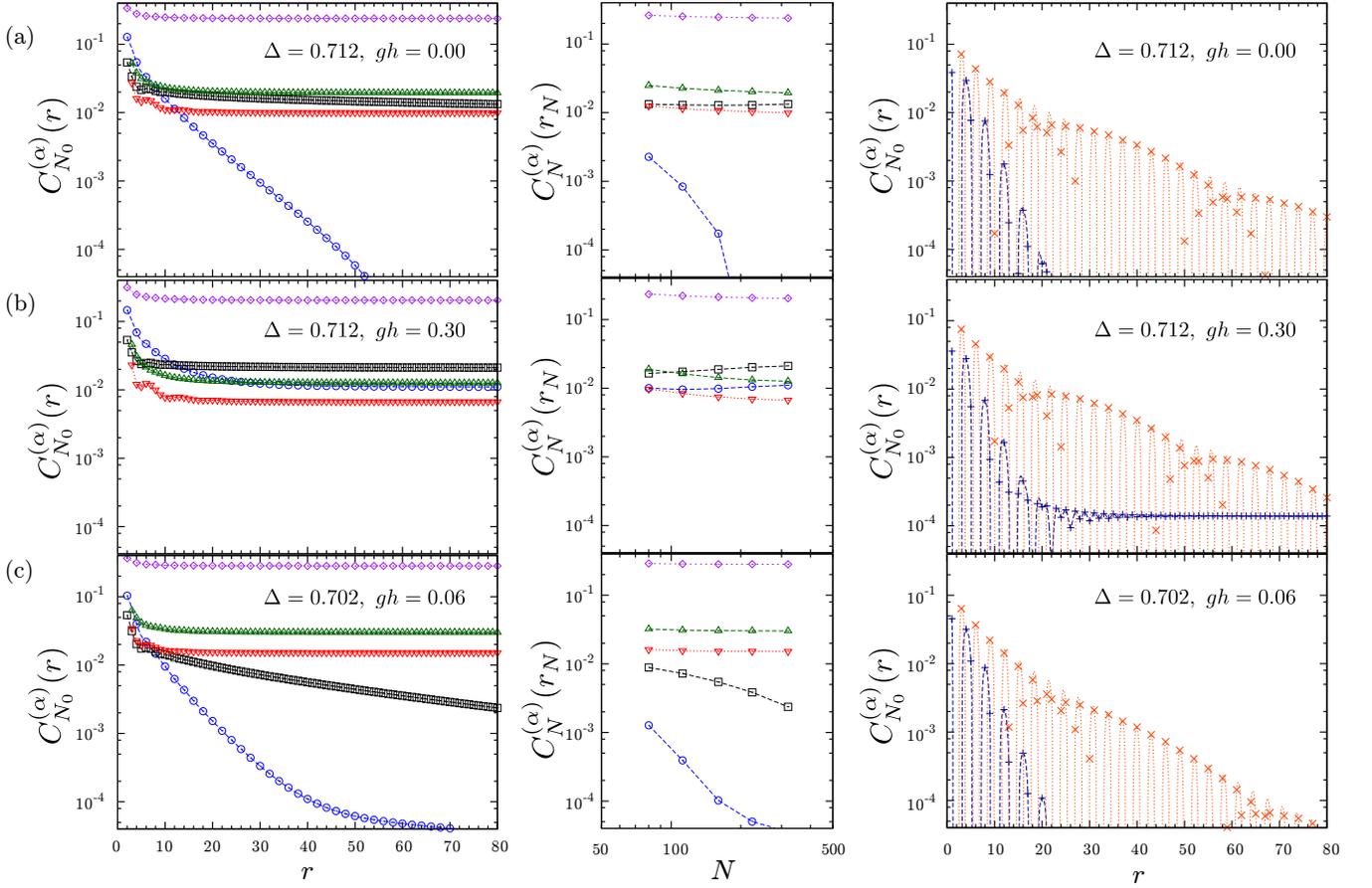}}
  \caption{ (Color online) 
Correlation functions $C^{(\alpha)}_{N}(r)$ for (a) $\Delta=0.712$, $gh=0.00$ in the VCD$_-$ phase, (b) $\Delta=0.712$, $gh=0.30$ in the VCND phase,  and (c) $\Delta=0.702$, $gh=0.06$ in the D$_-$ phase. The left and middle subpanels of (a)-(c) show the spatial decays of $C^{(\alpha)}_{N_0}(r)$ with $N_0=320$ and the $N$ dependence of $C^{(\alpha)}_{N}(r_N)$ with $r_N = N/4$ for $N=80, 112, 160, 224$ and $320$, respectively ($C^{(O^z_1)}_{N}(r):\textcolor{blue}{\bigcirc}$, $C^{(O^z_2)}_{N}(r):\textcolor{purple}{\Diamond}$, $C^{(\kappa^z)}_{N}(r):\Box$, $C^{(D^z)}_{N}(r):\textcolor{green}{\bigtriangleup}$, $C^{(D^x)}_{N}(r):\textcolor{red}{\bigtriangledown}$). The right subpanels show the spatial decay of spin correlation functions ($C^{(S^z)}_{N}(r):\textcolor{blue}{+}$, $C^{(S^x)}_{N}(r):\textcolor{red}{\times}$). The results are obtained by taking $m=300$.
}
\label{fig:vh_nonzero}
\end{figure*}

Now, we turn on the magnetic field $h$ along the $z$ direction. The leading $h$-linear term with respect to the number of spins involved is, of course, the Zeeman term $\hat{\mathcal{H}}_Z=-h\sum_j \hat{S}_j^z$ and the next-to-leading is a field-induced ring-exchange interaction~\cite{Sen95,Rokhsar90},
\begin{equation}
  \hat{ \mathcal{H} }_{\mathrm{ssc}} = gh \sum_{j} ( \hat{\bm{S}}_{ 2j - 1 } \! - \! \hat{\bm{S}}_{ 2j +2 } ) \cdot \hat{\bm{S}}_{2j} \times \hat{ \bm{S} }_{2j+1}.
  \label{eq:H_ssc}
\end{equation}
where $g$ is a linearized coupling constant proportional to a magnetic flux penetrating the triangle formed by three sites involved.
The sign of the flux alternates every other triangle, as shown in Fig.~\ref{fig_delta_vh}~(a), and hence it is proportional to the staggered scalar spin chirality.
When the ground state has an energy gap as in the D$_\pm$ and VCD$_\pm$ phases, however, $\hat{\mathcal{H}}_Z$ does not play any role at a much smaller Zeeman field than the energy gap. At such low field, only this staggered scalar chiral interaction $\hat{\mathcal{H}}_{\mathrm{ssc}}$ gives a nonvanishing contribution. In this respect, it is useful to consider its roles.

Now we tackle a phase diagram of $\hat{ \mathcal{H} }_{\delta\mathrm{xxz}} + \hat{ \mathcal{H} }_{\mathrm{ssc}}$ in the space of $\Delta$ and $gh$ near the zero-field ($h=0$) D$_-$-VCD$_-$ phase boundary $\Delta_{{\rm D}_- {\rm \mathchar`-VCD}_-} \simeq 0.709$ with other parameters being the same as in the previous sections, $J_1/J_2 = -2.5$ and $\delta=0.02$. The results are summarized in Fig.~\ref{fig_delta_vh} (b). The phase boundary has been determined in the same manner as in Fig.~\ref{fig:V=h=0}, namely, on a basis of long-range behaviors of the correlation functions as shown in Fig.~\ref{fig:vh_nonzero}. 

For a reference, we show in Fig.~\ref{fig:vh_nonzero}~(a) the correlation functions in the $gh=0$ case, namely, without $\hat{\mathcal{H}}{\mathrm{ssc}}$, for $\Delta=0.712$, which is located in the VCD$_-$ phase but nearly on the verge (Fig.~\ref{fig_delta_vh} (b)). While the VC and two dimer correlations are long-range from both the $r$ dependence with $N=320$ being fixed and the $N$ dependence at $r=r_N=N/4$, spin correlations clearly decay exponentially. However, as soon as we turn on $gh$, the $z$-component staggered spin correlation $C_N^{(S^z)}(r)$ grows and appears to become long-range at least for $gh=0.30$, as shown in Fig.~\ref{fig_delta_vh}~(b), indicating an emergence of the Neel LRO. It is also clear that the VC correlation $C^{\kappa^z}_{N}(r)$ is long-range. The dimer correlations $C^{D^\mu}_{N}(r)$ remain to be long-range and $\mathrm{Ave}_j\langle\hat{D}^x_j\rangle\mathrm{Ave}_j\langle\hat{D}^z_j\rangle<0$ holds (not shown). Thus, it is a coexistent phase of VC, Neel, and dimer orders, which is dubbed a VC Neel dimer (VCND) phase. 

Once the Neel LRO emergences, the both string correlations $\lim_{r\to\infty}C^{(O^z_1)}_{N}(2r)$ and $\lim_{r\to\infty}C^{(O^z_2)}_{N}(2r)$ become long-range, as shown in Fig.~\ref{fig_delta_vh}~(b). Actually, allowing for the Neel LRO, the two dimer D$_\pm$ phases are topologically connected. Namely, we cannot make any topological distinction from the sign of $\mathrm{Ave}_j\langle\hat{D}^x_j\rangle\mathrm{Ave}_j\langle\hat{D}^z_j\rangle$ within the VCND phase. This reflects that the absence of the Neel LRO is one of the symmetries that protect the distinction between the D$_+$ and D$_-$ phases in the sense of symmetry-protected topological phases. More precisely, this order breaks not only the time-reversal symmetry but also the invariance under the $D_2$ spin rotation about an axis in the $xy$ plane, which is required to having a nontrivial SPT in our model. A topological distinction will also be possible between the VCD$_+$ and VCD$_-$ phases which have a distinct nonzero string order parameter, as described by Eq.~\eqref{eq:string}. (At the moment, we do not know the local unitary transformation from one phase to the other and it is left for future study.) 

In the VCND phase, the Neel order parameter is extremely small and comparable to $10^{-2}$ for $gh=0.30$. It naturally becomes undetectably small for $gh<0.30$. Therefore, it is practically not possible to determine the VCD$_-$-VCND phase boundary from our numerics. We will see later in Sec.~\ref{sec:bosonization}, however, that the Neel order parameter is linearly coupled to the VC through the field-induced staggered scalar chiral interaction $\hat{ \mathcal{H} }_{\mathrm{ssc}}$ and thus induced by an infinitesimally small $gh$ if we start from the VCD$_\pm$ phase.
In effects, both of the time-reversal and the parity symmetries are spontaneously broken, but their product is fixed by $\hat{\mathcal{H}}_{\mathrm{ssc}}$. 

On the other hand, the D$_-$ phase is stable against the staggered scalar chiral interaction $\hat{ \mathcal{H} }_{\mathrm{ssc}}$, and it extends up to finite $gh$. Figure~\ref{fig:vh_nonzero}~(c) shows correlation functions for $gh=0.06$ and $\Delta=0.702$. The VC correlation and spin correlations all remain to be short-range. With further increasing $gh$, the system eventually enters the VCND phase, as shown in Fig.~\ref{fig_delta_vh}.

Finally, we briefly mention effects of $\hat{ \mathcal{H} }_{\mathrm{ssc}}$ around the D$_+$-VCD$_+$ phase boundary, i.e., $\Delta_{{\rm D}_+ {\rm \mathchar`-VCD}_+} \simeq 0.918$ for $J_1/J_2=-2.5$ and $\delta=0.02$. The VC LRO in the VCD$_+$ phase at $gh=0$ is stable at least up to a moderately large value of $gh$ but is also accompanied by the Neel LRO, realizing the VCND phase, as in the case of starting from the VCD$_-$ phase. 
However, even though we start from the D$_+$ phase and increase the coupling constant $gh$ of the staggered scalar chiral interaction $\hat{ \mathcal{H} }_{\mathrm{ssc}}$, the VC correlation does not become long-range (not shown), in contrast to the case of starting from the D$_-$ phase.

\section{Abelian bosonization analysis}
\label{sec:bosonization}
To gain a more insight on the gapped VC phases, we now perform an Abelian bosonization analysis of the total Hamiltonian $\hat{ \mathcal{H} }_{\delta\mathrm{xxz}} + \hat{ \mathcal{H} }_{\mathrm{ssc}} + \hat{ \mathcal{H} }_{Z}$. We start with the bosonized form of the Hamiltonian density for two antiferromagnetic ($J_2>0$) spin chains coupled in a zigzag manner with $J_1$ at zero Zeeman field but with a bond alternation $\delta>0$ and a finite staggered scalar chiral interaction $gh>0$,
\begin{align}
  \mathcal{H} &=\sum_{\nu=\pm}\frac{v_\nu}{2}\left[K_\nu(\partial_x\theta_\nu)^2+K_\nu^{-1}(\partial_x\phi_\nu)^2\right]
  \nonumber\\
  &
  -\gamma_1\cos\sqrt{4\pi}\phi_+\cos\sqrt{4\pi}\theta_-
  +\gamma_{\mathrm{tw}}(\partial_x\theta_+)\sin\sqrt{4\pi}\theta_-
  \nonumber\\
  &
  +\sum_{\nu=\pm}\gamma_{\mathrm{tw}}^{\nu,-\nu}(\partial_x\phi_\nu)\sin\sqrt{4\pi}\phi_{-\nu}
  \nonumber\\
  &
  -\gamma_\delta^{xy}\cos\sqrt{4\pi}\theta_--\sum_{\nu=\pm}\gamma_\delta^{z\nu}\cos\sqrt{4\pi}\phi_\nu
  \nonumber\\
  &
  -\gamma_{\mathrm{ssc}}(\partial_x\phi_-)\sin\sqrt{4\pi}\theta_--\gamma_{\mathrm{ssc}}'\sum_{\nu=\pm}\nu(\partial_x\phi_{-\nu})(\partial_x\theta_\nu).
\end{align}
We have introduced the bosonic fields $\theta_\pm=(\theta_1\pm\theta_2)/\sqrt{2}$ and $\phi_\pm=(\phi_1\pm\phi_2)/\sqrt{2}$ satisfying the commutation relation $[\phi_n(x),\theta_{n'}(x')]=i\delta_{n,n'}Y(x-x')$ with $Y(x)=0$ ($x<0$), 1/2 ($x=0$), and 1 ($x>0$). They are related to the spins through
\begin{align}
 &S^z_{2j+n}= \frac{a}{\sqrt{2\pi}} \partial_x \phi_n(x_n) + (-1)^j A_1 \cos (\sqrt{2\pi}\phi_n(x_n))+\dots,
 \label{eq:Sz}\\
 &S^+_{2j+n} 
 = e^{i\sqrt{2\pi}\theta_n(x_n)} \left[ (-1)^jB_0 + B_1 \cos(\sqrt{2\pi}\phi_n(x_n)) +\dots\right],
 \label{eq:Sp}
\end{align}
with $\Delta$-dependent non-universal constants, $A_1$, $B_0$, $B_1$ (Refs.~\onlinecite{Hikihara98,Lukyanov97}) and the lattice constant $a$. The velocities $v_\pm$ and the Tomonaga-Luttinger liquid parameters $K_\pm$ for the channel $\pm$ and the bare coupling constants $\gamma$'s take the following values;
\begin{align}
  &v_\pm\approx v\left(1\pm\frac{KJ_1\Delta a}{2\pi v}\right),
  ~~~
  v = \frac{\pi\sqrt{1-\Delta^2}}{2\arccos\Delta}J_2a,
  \notag\\
  &K_\pm\approx K\left(1\mp\frac{KJ_1\Delta a}{2\pi v}\right),
  ~~~
  K = \left[1-(1/\pi)\arccos\Delta\right]^{-1},
  \notag\\
  &\gamma_1=B_1^2J_1/a,
  ~~
  \gamma_{\mathrm{tw}}=\sqrt{\pi}J_1B_0^2,
  ~~
  \gamma_{\mathrm{tw}}^{\pm,\mp}=\sqrt{\pi}A_1^2J_1\Delta/2,
  \notag\\
  &\gamma_\delta^{xy}=-2B_0^2J_1\delta/a,
  ~~
  \gamma_\delta^{z\pm}=-A_1^2J_1 \Delta \delta/a,
  \notag\\
  &\gamma_{\mathrm{ssc}}=2\sqrt{\pi}B_0^2gh,
  ~~\gamma_{\mathrm{ssc}}'=(B_0^2+B_1^2/2)agh.
\end{align}
Note the discrepancy from the result on $\hat{ \mathcal{H} }_{\mathrm{ssc}}$ in Ref.~\onlinecite{Sen95}, which included a relevant symmetry-forbidden term simply proportional to $\sin\sqrt{4\pi}\theta_-$.

It has been shown that in the case of $\delta=0$ ($\gamma_\delta^{xy}=\gamma_\delta^{z\pm}=0$) and $g=0$ ($\gamma_{\mathrm{ssc}}=\gamma_{\mathrm{ssc}}'=0$), the gapless chiral phase can appear as the coupling constant $\gamma_{\mathrm{tw}}$ grows on the renormalization group flow for a certain range of $J_1/J_2<0$ and $0<\Delta<1$ [Refs.~\onlinecite{Nersesyan98,FSOF12}]. The bond alternation of $J_1$ immediately locks the field $\sqrt{4\pi}\phi_+$ to either 0 or $\pi$ depending on the sign of $J_1\delta$ and uniformly shifts the locking position of $\sqrt{4\pi}\theta_-$ from $\pm\pi/2$, since it provides relevant interactions with the coupling constants $\gamma_\delta^{xy}$ and $\gamma_\delta^{z+}$. This yields a coexistence of the vector-chiral and dimer orders characterized by $\langle\sin\sqrt{4\pi}\theta_-\rangle\ne0$ and $(\langle\cos\sqrt{4\pi}\theta_-\rangle,\langle\cos\sqrt{4\pi}\phi_+\rangle)\ne0$, respectively, while the Neel LRO is absent, i.e., $\langle\partial_x\phi_-\rangle=\langle\sin\sqrt{4\pi}\phi_+\rangle=0$. As far as the fields are locked, we can rely on a mean-field approximation to decouple the $\pm$ sectors~\cite{Nersesyan98,FSOF12}. Then, the Hamiltonian density is written as $\mathcal{H}\approx\mathcal{H}_0+\mathcal{H}_++\mathcal{H}_-$ with
\begin{align}
  \mathcal{H}_0
  &=\gamma_1\langle\cos\sqrt{4\pi}\phi_+\rangle\langle\cos\sqrt{4\pi}\theta_-\rangle
  +\gamma_{\mathrm{tw}}Q_{\theta_+}\langle\sin\sqrt{4\pi}\theta_-\rangle
  \nonumber\\
  &+\gamma_{\mathrm{tw}}^{-+}Q_{\phi_-}\langle\sin\sqrt{4\pi}\phi_+\rangle
  \nonumber\\
  &-\left[\gamma_{\mathrm{ssc}}\langle\sin\sqrt{4\pi}\theta_-\rangle-\gamma_{\mathrm{ssc}}'Q_{\theta_+}\right]Q_{\phi_-},
  \label{eq:H_0}\\
  \mathcal{H}_+
  &=\frac{v_+}{2}\left[K_+((\partial_x\tilde{\theta}_+)^2-Q_{\theta_+}^2)+K_+^{-1}(\partial_x\phi_+)^2\right]
  \nonumber\\
  &-(\gamma_\delta^{z+}+\gamma_1\langle\cos\sqrt{4\pi}\theta_-\rangle)\cos\sqrt{4\pi}\phi_+
  \nonumber\\
  &-\gamma_{\mathrm{tw}}^{-+}Q_{\phi_-}\sin\sqrt{4\pi}\phi_+,
  \label{eq:H_+}\\
  \mathcal{H}_-
  &=\frac{v_-}{2}\left[K_-(\partial_x\theta_-)^2+K_-^{-1}((\partial_x\tilde{\phi}_-)^2-Q_{\phi_-}^2)\right]
  \nonumber\\
  &-(\gamma_\delta^{xy}+\gamma_1\langle\cos\sqrt{4\pi}\phi_+\rangle)\cos\sqrt{4\pi}\theta_-
  \nonumber\\
  &-(\gamma_{\mathrm{tw}}Q_{\theta_+}-\gamma_{\mathrm{ssc}}Q_{\phi_-})\sin\sqrt{4\pi}\theta_-,
  \label{eq:H_-}
\end{align}
with $\tilde{\theta}_+=\theta_++Q_{\theta_+}x$ and $\tilde{\phi}_-=\phi_-+Q_{\phi_-}x$, where
\begin{align}
  Q_{\theta_+}
  &=\frac{\gamma_{\mathrm{tw}}v_--\gamma_{\mathrm{ssc}}\gamma_{\mathrm{ssc}}'K_-}{v_+v_-K_+-\gamma_{\mathrm{ssc}}'^2K_-}
  \langle\sin\sqrt{4\pi}\theta_-\rangle
  \notag\\
  &+\frac{\gamma_{\mathrm{ssc}}'K_-\gamma_{\mathrm{tw}}^{-+}}{v_+v_-K_+-\gamma_{\mathrm{ssc}}'^2K_-}
  \langle\sin\sqrt{4\pi}\phi_+\rangle,
  \label{eq:Q_theta+}\\
  Q_{\phi_-}
  &=\frac{K_-}{v_-}\left[\frac{\gamma_{\mathrm{tw}}\gamma_{\mathrm{ssc}}'-\gamma_{\mathrm{ssc}}v_+K_+}{v_+K_+-\gamma_{\mathrm{ssc}}'^2K_-/v_-}
    \langle\sin\sqrt{4\pi}\theta_-\rangle
    \right.\notag\\
    &\left.~~~~~~~
    +\frac{v_+K_+\gamma_{\mathrm{tw}}^{-+}}{v_+K_+-\gamma_{\mathrm{ssc}}'^2K_-/v_-}
    \langle\sin\sqrt{4\pi}\phi_+\rangle\right].
  \label{eq:Q_phi-}
\end{align}
Note that $\mathcal{H}_\pm$ can be solved to give
\begin{widetext}
\begin{align}
  \left\{\begin{array}{c}\langle\sin\sqrt{4\pi}\phi_+\rangle\\
  \langle\cos\sqrt{4\pi}\phi_+\rangle
  \end{array}\right\}
  &=\frac{C(K_+)}{v_+/a^2}\left\{\begin{array}{c}\gamma_{\mathrm{tw}}^{-+}Q_{\phi_-}\\
  (\gamma_\delta^{z+}+\gamma_1\langle\cos\sqrt{4\pi}\theta_-\rangle)
  \end{array}\right\}
  \left(\frac{a^2}{v_+}\sqrt{(\gamma_\delta^{z+}+\gamma_1\langle\cos\sqrt{4\pi}\theta_-\rangle)^2+(\gamma_{\mathrm{tw}}^{-+})^2Q_{\phi_-}^2}\right)^{-\frac{2(K_+^{-1}-1)}{2K_+^{-1}-1}},
  \label{eq:sc_phi+}\\
  \left\{\begin{array}{c}\langle\sin\sqrt{4\pi}\theta_-\rangle\\
  \langle\cos\sqrt{4\pi}\theta_-\rangle
  \end{array}\right\}
  &=\frac{C(K_-^{-1})}{v_-/a^2}\left\{\begin{array}{c}\gamma_{\mathrm{tw}}Q_{\theta_+}-\gamma_{\mathrm{ssc}}Q_{\phi_-}\\
  (\gamma_\delta^{xy}+\gamma_1\langle\cos\sqrt{4\pi}\phi_+\rangle)
  \end{array}\right\}
  \left(\frac{a^2}{v_-}\sqrt{(\gamma_\delta^{xy}+\gamma_1\langle\cos\sqrt{4\pi}\phi_+\rangle)^2+(\gamma_{\mathrm{tw}}Q_{\theta_+}-\gamma_{\mathrm{ssc}}Q_{\phi_-})^2}\right)^{-\frac{2(K_--1)}{2K_--1}},
  \label{eq:sc_theta-}
\end{align}
\end{widetext}
with $C(\beta)$ being the nonuniversal constant in the solution to the sine-Gordon model~\cite{Lukyanov97}, where we have exploited the fact that the locking positions $\langle\phi_+\rangle$ and $\langle\theta_-\rangle$ of $\phi_+$ and $\theta_-$, respectively, are given through
\begin{align}
  \sin\sqrt{4\pi}\langle\phi_+\rangle&=F\left(\gamma_{\mathrm{tw}}^{-+}Q_{\phi_-},\gamma_\delta^{z+}+\gamma_1\langle\cos\sqrt{4\pi}\theta_-\rangle\right),
  \notag\\
  \cos\sqrt{4\pi}\langle\phi_+\rangle&=F\left(\gamma_\delta^{z+}+\gamma_1\langle\cos\sqrt{4\pi}\theta_-\rangle,\gamma_{\mathrm{tw}}^{-+}Q_{\phi_-}\right),
  \notag\\
  \sin\sqrt{4\pi}\langle\theta_-\rangle&=F\left(\gamma_{\mathrm{tw}}Q_{\theta_+}-\gamma_{\mathrm{ssc}}Q_{\phi_-},\gamma_\delta^{xy}+\gamma_1\langle\cos\sqrt{4\pi}\phi_+\rangle\right),
  \notag\\
  \cos\sqrt{4\pi}\langle\theta_-\rangle&=F\left(\gamma_\delta^{xy}+\gamma_1\langle\cos\sqrt{4\pi}\phi_+\rangle,\gamma_{\mathrm{tw}}Q_{\theta_+}-\gamma_{\mathrm{ssc}}Q_{\phi_-}\right),
  \label{eq:dphi_+,dtheta_-}
\end{align}
with $F(x,y)=x/\sqrt{x^2+y^2}$. Equations~\eqref{eq:Q_theta+}, \eqref{eq:Q_phi-}, \eqref{eq:sc_phi+}, and \eqref{eq:sc_theta-} are solved self-consistently to get the solution within the mean-field decoupling approximation of $(\theta_+,\phi_+)$ and $(\theta_-,\phi_-)$ modes.

\subsection{A transition between VCD$_+$ and VCD$_-$ phases}
Let us consider the case without the staggered scalar chiral interaction, i.e., $\gamma_{\mathrm{ssc}}=\gamma_{\mathrm{ssc}}'=0$. Then, introducing a function $\Theta(y)=y$ for $y\in[-1,1]$ and $\sgn(y)$ otherwise, we obtain 
\begin{widetext}
\begin{align}
  \cos\sqrt{4\pi}\langle\phi_+\rangle_0
  &=\Theta\left(\frac{\gamma_\delta^{z+}+\gamma_1\langle\cos\sqrt{4\pi}\theta_-\rangle_0}{v_+/a^2}\left[C(K_+)K_-\frac{(\gamma_{\mathrm{tw}}^{-+})^2}{v_+v_-/a^2}\right]^{-\frac{2K_+^{-1}-1}{2(K_+^{-1}-1)}}\right),
  \notag\\
  \frac{Q_{\phi_-}}{\gamma_{\mathrm{tw}}^{-+}}
  =\langle\sin\sqrt{4\pi}\phi_+\rangle_0
  &=\pm C(K_+)\left[C(K_+)K_-\frac{(\gamma_{\mathrm{tw}}^{-+})^2}{v_+v_-/a^2}\right]^{\frac{1}{2(K_+^{-1}-1)}}\sqrt{1-\cos^2\sqrt{4\pi}\langle\phi_+\rangle_0},
  \notag\\
  \langle\cos\sqrt{4\pi}\phi_+\rangle_0
  &=C(K_+)\left(\frac{\gamma_\delta^{z+}+\gamma_1\langle\cos\sqrt{4\pi}\theta_-\rangle_0}{v_+/a^2}\right)^{\frac{1}{2K_+^{-1}-1}}\left[\cos\sqrt{4\pi}\langle\phi_+\rangle_0\right]^{1-\frac{1}{2K_+^{-1}-1}},
  \notag\\
  \cos\sqrt{4\pi}\langle\theta_-\rangle_0
  &=\Theta\left(\frac{\gamma_\delta^{xy}+\gamma_1\langle\cos\sqrt{4\pi}\phi_+\rangle_0}{v_-/a^2}\left[C(K_-^{-1})K_+^{-1}\frac{\gamma_{\mathrm{tw}}^2}{v_+v_-/a^2}\right]^{-\frac{2K_--1}{2(K_--1)}}\right),
  \notag\\
  \frac{Q_{\theta_-}}{\gamma_{\mathrm{tw}}/v_+K_+}=\langle\sin\sqrt{4\pi}\theta_-\rangle_0
  &=\pm C(K_-^{-1})\left[C(K_-^{-1})K_+^{-1}\frac{\gamma_{\mathrm{tw}}^2}{v_+v_-/a^2}\right]^{\frac{1}{2(K_--1)}}\sqrt{1-\cos^2\sqrt{4\pi}\langle\theta_-\rangle_0},
  \notag\\
  \langle\cos\sqrt{4\pi}\theta_-\rangle_0
  &=C(K_-^{-1})\left(\frac{\gamma_\delta^{xy}+\gamma_1\langle\cos\sqrt{4\pi}\phi_+\rangle_0}{v_-/a^2}\right)^{\frac{1}{2K_--1}}\left[\cos\sqrt{4\pi}\langle\theta_-\rangle_0\right]^{1-\frac{1}{2K_--1}}.
  \label{eq:sol}
\end{align}
\end{widetext}
where $\langle\cdots\rangle_0$ denotes the average taken for $g=0$.

Suppose the system is in a gapped VC dimer (VCD$_\pm$) state, characterized by the absence of the Neel order, i.e., $Q_{\phi_-}=\langle\sin\sqrt{4\pi}\phi_+\rangle_0=0$, the presence of the VC LRO, i.e., $\langle\sin\sqrt{4\pi}\theta_-\rangle_0>0$, and the presence of the $z$-component dimer order, i.e., $\langle\cos\sqrt{4\pi}\phi_+\rangle_0\ne0$. Then, it requires the conditions
\begin{align}
  &\left|\frac{\gamma_\delta^{z+}+\gamma_1\langle\cos\sqrt{4\pi}\theta_-\rangle_0}{v_+/a^2}\right|\ge\left[C(K_+)K_-\frac{(\gamma_{\mathrm{tw}}^{-+})^2}{v_+v_-/a^2}\right]^{\frac{2K_+^{-1}-1}{2(K_+^{-1}-1)}},
  \label{eq:VCD:s_phi+=0}\\
  &\frac{\gamma_\delta^{xy}+\gamma_1\langle\cos\sqrt{4\pi}\phi_+\rangle_0}{v_-/a^2}<\left[C(K_-^{-1})K_+^{-1}\frac{\gamma_{\mathrm{tw}}^2}{v_+v_-/a^2}\right]^{\frac{2K_--1}{2(K_--1)}},
  \label{eq:VCD:s_theta->0}\\
  &K_+<1, ~~~ K_->1.
  \label{eq:VCD:K}
\end{align}
Furthermore, Eqs.~\eqref{eq:sol} are reduced to
\begin{align}
  \langle\cos\sqrt{4\pi}\phi_+\rangle_0
  &=C(K_+)\left(\frac{\gamma_\delta^{z+}+\gamma_1\langle\cos\sqrt{4\pi}\theta_-\rangle_0}{v_+/a^2}\right)^{\frac{1}{2K_+^{-1}-1}},
  \label{eq:VCD:c_phi+}\\
  \langle\cos\sqrt{4\pi}\theta_-\rangle_0
  &=\frac{v_+K_+}{\gamma_{\mathrm{tw}}^2}(\gamma_\delta^{xy}+\gamma_1\langle\cos\sqrt{4\pi}\phi_+\rangle_0),
  \label{eq:VCD:c_theta-}\\
  \langle\sin\sqrt{4\pi}\phi_+\rangle_0&=0,
  \label{eq:VCD:s_phi+}
\end{align}
\begin{widetext}
\begin{equation}
  \langle\sin\sqrt{4\pi}\theta_-\rangle_0
  =\pm C(K_-^{-1})\sqrt{\left[\frac{v_+K_+}{C(K_-^{-1})\gamma_{\mathrm{tw}}^2}\frac{v_-}{a^2}\right]^{\frac{1}{1-K_-}}-\left(\frac{v_+K_+}{C(K_-^{-1})\gamma_{\mathrm{tw}}^2}(\gamma_\delta^{xy}+\gamma_1\langle\cos\sqrt{4\pi}\phi_+\rangle_0)\right)^2}.
  \label{eq:VCD:s_theta-}
\end{equation}
\end{widetext}
The values of $\langle\cos\sqrt{4\pi}\phi_+\rangle$ and $\langle\cos\sqrt{4\pi}\theta_-\rangle$ are obtained as the self-consistent solution to Eqs.~\eqref{eq:VCD:c_phi+} and \eqref{eq:VCD:c_theta-}.

Now, let us consider a transition from VCD$_+$ to VCD$_-$, at which $\langle\cos\sqrt{4\pi}\phi_+\rangle$ changes its sign. If we assume it is a continuous transition, then $D^z=\mathrm{Ave}_j\langle\hat{D}^z_j\rangle$ and thus $\langle\cos\sqrt{4\pi}\phi_+\rangle$ must vanish at the transition, which occurs as $\gamma_\delta^{z+}+\gamma_1\langle\cos\sqrt{4\pi}\theta_-\rangle\to0$. However, this violates the condition for $\langle\sin\sqrt{4\pi}\phi_+\rangle=0$, namely, Eq.~\eqref{eq:VCD:s_phi+=0}, unless $K_+$ accidentally decays to unity in such a way that Eq.~\eqref{eq:VCD:s_phi+=0} is always satisfied. Then, the following four possible scenarios arise: If this accidental condition is satisfied, then VCD$_+$ and VCD$_-$ phases are separated either (i) by a direct continuous transition at a single point or, more in general, (ii) by an extended area of a gapless VCD$_0$ phase, where the system becomes gapless since $\phi_+$ is unlocked. Otherwise, (iii) it may happen to induce the Neel order with $N^z=\mathrm{Ave}_j(-1)^j\langle\hat{S}^z_j\rangle\propto\langle\sin\sqrt{4\pi}\phi_+\rangle\ne0$ at a single transition point where the locking position $\sqrt{4\pi}\langle\phi_+\rangle$ discontinuously jumps to $\pi/2$, leading to $N^z\ne0$ and $D^z=0$. Or more in general, (iv) the Neel dimer state extends to a finite region over which $\sqrt{4\pi}\langle\phi_+\rangle$ continuously changes from 0 to $\pi$ or from $\pi$ to 0, and the system remains gapped.

Our numerical calculations shown in Sec.~\ref{sec:0mag} indicates that around the VCD$_+$-VCD$_-$ transition, i.e., around $\Delta=0.88$ for $J_1/J_2=-2.5$ and $\delta=0.02$, the $z$-component dimer order parameter $D^z$ vanishes and changes the sign while the VC order and the inplane dimer order remain finite. The correlation length for $C_N^{(S^x)}(r)$ grows longer than half the system size accordingly, while that for $C_N^{(S^z)}(r)$ decreases with decreasing $\Delta$. Hence, the system is most likely gapless without the Neel order in this region. This supports the scenario (i) or (ii), although it is difficult to judge which the case is. At least, the transition region is restricted to a quite narrow region, narrower than $0.879\le\Delta\le0.883$, as shown in Sec.~\ref{sec:0mag:entanglement}.

\subsection{Effects of the staggered scalar chiral interaction on the vector-chiral dimer phases}
The effect of the staggered scalar chiral interaction, $gh>0$, on the VC dimer phases is easily understood as follows:
 We first assume that the conditions of Eqs.~\eqref{eq:VCD:s_phi+=0}, \eqref{eq:VCD:s_theta->0}, and \eqref{eq:VCD:K} hold at least for $gh=0$. Then, we expand the self-consistent equations of the mean-field decoupling approximation, Eqs.~\eqref{eq:Q_theta+}, \eqref{eq:Q_phi-}, \eqref{eq:sc_phi+}, and \eqref{eq:sc_theta-}, up to the linear order in $g$ (thus in $\gamma_{\mathrm{ssc}}$ and $\gamma_{\mathrm{ssc}}'$) around the solution at $g=0$, Eqs.~\eqref{eq:VCD:c_phi+}, \eqref{eq:VCD:c_theta-}, \eqref{eq:VCD:s_phi+}, and \eqref{eq:VCD:s_theta-}. In particular, Eqs.~\eqref{eq:Q_theta+} and \eqref{eq:Q_phi-} are written as
\begin{align}
  Q_{\theta_+}
  &=\frac{\gamma_{\mathrm{tw}}}{v_+K_+}\left[
    \langle\sin\sqrt{4\pi}\theta_-\rangle_0+\langle\sin\sqrt{4\pi}\theta_-\rangle_1\right]
  +\cdots,
  \label{eq:exp:Q_theta+}\\
  Q_{\phi_-}
  &=\frac{K_-}{v_-}\left(\frac{\gamma_{\mathrm{tw}}\gamma_{\mathrm{ssc}}'}{v_+K_+}-\gamma_{\mathrm{ssc}}\right)
    \langle\sin\sqrt{4\pi}\theta_-\rangle_0
    \notag\\
    &~~~
    +\frac{\gamma_{\mathrm{tw}}^{-+}K_-}{v_-}\langle\sin\sqrt{4\pi}\phi_+\rangle_1+\cdots,
  \label{eq:exp:Q_phi-}
\end{align}
where $\langle\cdots\rangle_1$ denotes the average in the linear order in $g$. $\langle\sin\sqrt{4\pi}\phi_+\rangle_1$ can be obtained as the following nonvanishing expression by plugging Eq.~\eqref{eq:exp:Q_phi-} into Eq.~\eqref{eq:sc_phi+} and subsequently expanding the equation in terms of $gh$;
\begin{widetext}
\begin{align}
  \langle\sin\sqrt{4\pi}\phi_+\rangle_1
  &=
  \left(\frac{\gamma_{\mathrm{tw}}\gamma_{\mathrm{ssc}}'}{v_+K_+}-\gamma_{\mathrm{ssc}}\right)
  \langle\sin\sqrt{4\pi}\theta_-\rangle_0
  \left[\frac{v_+v_-}{C(K_+)K_-\gamma_{\mathrm{tw}}^{-+}a^2}\left(\frac{a^2}{v_+}\left|\gamma_\delta^{z+}+\gamma_1\langle\cos\sqrt{4\pi}\theta_-\rangle\right|\right)^{\frac{2(K_+^{-1}-1)}{2K_+^{-1}-1}}-\gamma_{\mathrm{tw}}^{-+}\right]^{-1}.
\end{align}
\end{widetext}
Note that $N^z\propto Q_{\phi_-}\propto\langle\sin\sqrt{4\pi}\phi_+\rangle$. This confirms that the Neel order is induced by turning on an infinitesimally small coupling constant $gh$ of the staggered scalar chiral interaction if we start from the VC dimer (VCD$_\pm$) phases.

\section{The case at moderately large magnetic field: Zeeman interaction}
\label{sec:Zeeman}
 \begin{figure}[H!tb]
   \centering
   \resizebox{8.5cm}{!}{\includegraphics{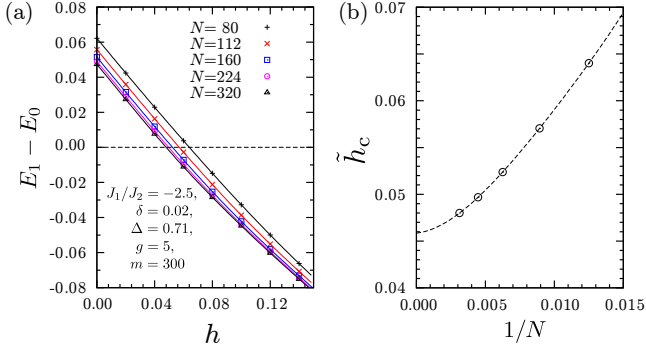}}
   \caption{ (Color online) 
 (a) Energy difference between lowest-energy of $S^z_{\rm tot} = 1$ ($E_1$) and   $S^z_{\rm tot} = 0$ ($E_0$). Interpolation curves of $E_1 - E_0$ as function of $h$ are the cubic spline. The level cross point, namely $E_1 - E_0 = 0$, is represented by $\tilde{h}_{\rm c}$. (b) Finite-size dependence of $\tilde{h}_{\rm c}$. The finite-size data are naively extrapolated to $h_c \simeq 0.0459$ as $N \rightarrow \infty$ using the polynomial function. 
 }
 \label{fig:cross_point}
 \end{figure}
\begin{figure}[H!tb]
  \centering
  \resizebox{8.0cm}{!}{\includegraphics{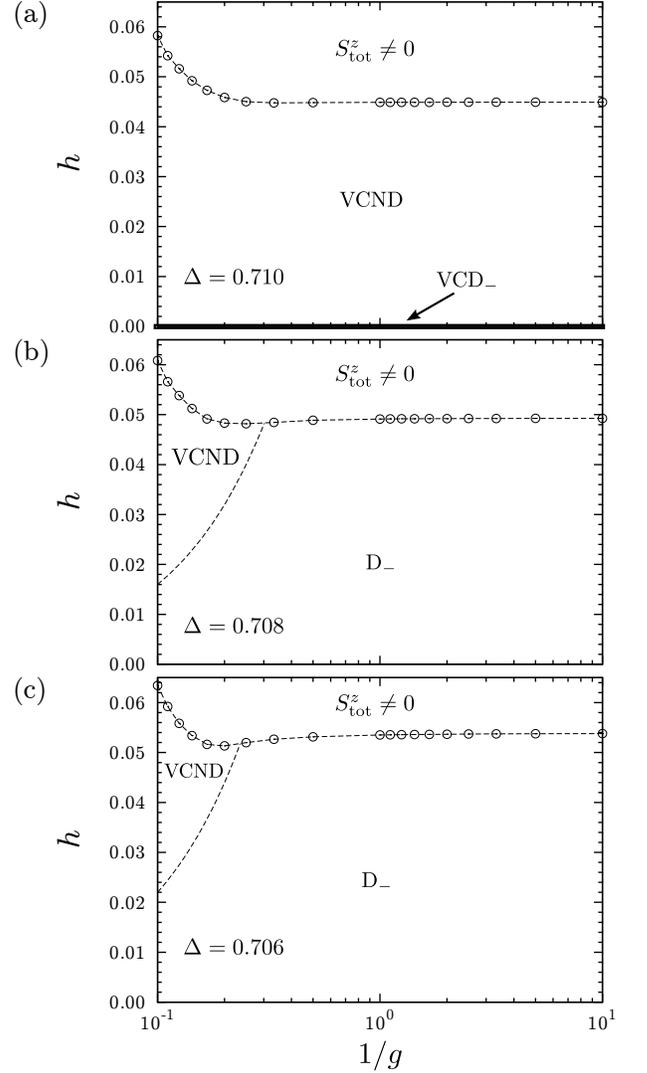}}
  \caption{ 
Phase diagram of low magnetic field region for $\hat{ \mathcal{H} }_{\delta\mathrm{xxz}} + \hat{ \mathcal{H} }_{\mathrm{ssc}} + \hat{ \mathcal{H} }_{Z}$ at $J_1/J_2 = -2.5$, $\delta=0.02$: (a) $\Delta = 0.71$, (b) $\Delta = 0.708$, and (c) $\Delta = 0.706$.  The broken lines between VCD$_-$ and D$_-$ in (b) and (c) represent $gh = 0.16$ and $0.22$, respectively, which are the transition values as shown in Fig.~\ref{fig_delta_vh}. 
The number of retain state $m$ in the DMRG method is 300. 
}
\label{fig:h_1ov_D708}
\end{figure}

At a moderately large magnetic field, the Zeeman field can be comparable to or larger than the zero-field energy gap and induce a finite magnetization. Then, we need to consider the total Hamiltonian $\hat{ \mathcal{H} }_{\delta\mathrm{xxz}} + \hat{ \mathcal{H} }_{\mathrm{ssc}} + \hat{ \mathcal{H} }_{Z}$. In particular, we focus on the parameter region in the vicinity of the zero-field D$_-$-VCD$_-$ phase boundary, i.e., $\Delta_{ {\rm D}_- {\rm \mathchar`-VCD}_-} \simeq 0.709$ for $J_1/J_2=-2.5$ and $\delta=0.02$ (see Fig.~\ref{fig:scaling} (a)). 
The results on the phase diagram are summarized in Fig.~\ref{fig:h_1ov_D708}. The emergence of a uniform magnetization has been probed by carefully examining a level crossing of $E_0$ and $E_1$ as a function of the field $h$, where $E_n$ represents the lowest energy level with the eigenvalue $\hat{S}^z_{\mathrm{tot}}=n$. Namely, for a given value of $g$, we obtain as a function of $N$ the threshold field value $\tilde{h}_c$ beyond which the energy level difference $E_1-E_0$ becomes negative, and subsequently extrapolate $\tilde{h}_c$ into $N\to\infty$, as exemplified in Fig.~\ref{fig:cross_point}~(a) and (b), respectively. The critical value $h_c$ roughly corresponds to the energy gap of the zero-field ground state. (See also Fig.~\ref{fig:cross_point}~(a)). Note that in the region we studied, either $E_0$ or $E_1$ gives the ground-state energy. 

If we apply the magnetic field to the zero-field VCD$_-$ ground state, namely, for $\Delta_{{\rm D}_- {\rm \mathchar`-VCD}_-} \simeq 0.709$ in the case of $J_1/J_2=-2.5$ and $\delta=0.02$, it acquires the Neel LRO as soon as we turn on the field, because of the staggered scalar chiral interaction, as already shown in Sec.~\ref{sec:bosonization}. It yields the VCND phase at low field $h$. With increasing field $h$, the $\hat{S}^z_{\mathrm{tot}}$ eigenvalue of the ground state changes from 0 to 1, as shown for $\Delta=0.710$ in Fig.~\ref{fig:h_1ov_D708}~(a), signaling an onset of a finite uniform magnetization.

On the other hand, the D$_-$ ground state is stable up to a certain magnetic field shown in Fig.~\ref{fig:h_1ov_D708}~(b) and (c) for $\Delta=0.708$ and 0.706, respectively. The VCND phase also appears in the phase diagram but only at a finite magnetic field $h$ for large $g$. Decreasing $\Delta$ from the zero-field D$_-$-VCD$_-$ phase boundary $\Delta_{{\rm D}_- {\rm \mathchar`-VCD}_-}$ narrows the VCND phase in the space of $h$ and $1/g$.

We have also performed DMRG calculations for $\Delta>\Delta_{{\rm D}_+ {\rm \mathchar`-VCD}_+}$ for the same set of parameters $J_1/J_2=-2.5$ and $\delta=0.02$, which fall into the Haldane dimer (D$_+$) phase. However, in any $\hat{S}^z_{\mathrm{tot}}$ ground state, the VC-Neel LRO induced by the staggered scalar chiral interaction does not appear.

Ground-state properties of the phase with a finite uniform magnetization are beyond the scope of this paper. It is most natural that it is a gapless VCND$_0$ phase where the $x,y$-component dimer order parameter remains finite but the $z$-component vanishes, and that the Neel order parameter vanishes in the limit of $v\to0$. Then, the phase is characterized in terms of the following properties of the bosonized fields defined in Sec.~\ref{sec:bosonization}: The $\theta_-$ field is locked with an incommensurate value of $\langle\sqrt{4\pi}\theta_-\rangle$ so that both $\mathrm{Ave}_j\langle\kappa^z_{j+1/2}\rangle\propto \langle\sin\sqrt{4\pi}\theta_-\rangle$ and $\mathrm{Ave}_j\langle\hat{D}^x_{j}\rangle\propto \langle\cos\sqrt{4\pi}\theta_-\rangle$ are nonzero. The $(\theta_+,\phi_+)$ sector is unlocked with finite values of the magnetization $\langle\hat{S}^z_{\mathrm{tot}}\rangle\propto\langle\partial_x\phi_+\rangle$ and the incommensurability $\langle\partial\theta_+\rangle$ of a quasi-LRO of inplane spin correlations.

\section{conclusions and remarks}
\label{sec:conclustion}

We have investigated a $J_1$-$J_2$ frustrated XXZ spin-1/2 chain with the nearest-neighbor ferromagnetic exchange coupling alternating in its magnitude as $J_1(1\pm\delta)$. The zero-field phase diagram in this case is modified from the $\delta=0$ case as follows. (See Fig.~\ref{fig:V=h=0}.)

The vector-chiral (VC) LRO is reduced by a small $\delta$, though it does not completely disappear. In particular, we have observed a disappearance of the gapless VC phase, which appears in a wide range of parameters $\Delta$ and $J_1/J_2$ ($J_1<0$, $J_2>0$) for $\delta=0$ and should eventually evolve into a spin-spiral magnetic LRO in three dimensions. It appears either at a single point or at most only in a quite narrow region, which lies between two gapped VC dimer VCD$_+$ and VCD$_-$ phases. Here the subscript $\pm$ denotes the relative sign of $x,y$- and $z$-component dimer order parameters. The gapless VC dimer VCD$_0$ state appears only in a region where the $z$-component dimer order happens to vanish upon the sign reversal. Increasing $\Delta$ from the VCD$_+$ phase towards the SU(2)-symmetric case $\Delta=1$, it undergoes a phase transition beyond which the VC order parameter disappears and the system is described as a Haldane dimer (D$_+$) phase. Similarly, decreasing $\Delta$ from the VCD$_-$ phase towards the U(1)-symmetric case $\Delta=0$, the VC order also eventually vanishes and the system enters the even-parity dimer (D$_-$) phase.

A large enough Zeeman field to close the spin gap induces the gapless vector-chiral phase, as in the $\delta=0$ case~\cite{Kolezhuk05, Hikihara08, FSOF12}. However, the Zeeman term does not play any role at smaller field strength as long as the magnetization vanishes. In this case, the staggered scalar chiral interaction $\hat{ \mathcal{H} }_{\mathrm{ssc}}$, namely, Eq.~(\ref{eq:H_ssc}), due to the magnetic flux penetrating triangles formed by three nearby spins~\cite{Sen95,Rokhsar90}, is of primary importance. It can actually induce a coexistent phase of the VC, Neel, and $x,y$- and $z$-component dimer orders, i.e., the VCND phase. This coexistence is readily realized when we start from the VCD$_-$ phase or from the D$_-$ phase near the D$_-$-VCD$_-$ phase boundary. On the other hand, if we start from the D$_+$ phase on a more SU(2)-symmetric side, we do not observe the VCND phase.

Finally, we address a possible relevance to experiments. As we have mentioned in sec.~\ref{sec:intro}, it has been found that Rb$_2$Cu$_2$Mo$_3$O$_{12}$ provides a unique case among quasi-one-dimensional (Q1D) frustrated spin-1/2 systems based on cuprates. It does not exhibit a spin-spiral LRO and the associated ferroelectricity at zero magnetic field, unlike many Q1D cuprates, but has an energy gap, of the order of 0.2~meV, in spin excitations~\cite{exp}. Applying the magnetic field, it does show a ferroelectricity although the spin excitations remain gapped with nearly zero magnetization without a clear signal of spin-spiral LRO. These novel experimental features can be understood in terms of our theory, at least, at a qualitative level: First of all, there exists a crystallographic dimerization in this compound~\cite{Solodovnikov97}. It is likely that at lower temperatures than the spin gap, the magnitude of the dimerization is enhanced by correlations, so that our choice of $\delta=0.02$ might be within a practical range.
Then, with easy-plane exchange anisotropy, the system is readily located in the D$_-$ phase but in a vicinity of the D$_-$-VCD$_-$ phase boundary at zero magnetic field. Therefore it is gapped and shows no long-range or even quasi-long-range spin correlations, in agreements with experiments. Turning on the magnetic field, the staggered scalar chiral interaction first induces the VC LRO accompanied by a Neel LRO whose amplitude might be too tiny to detect experimentally. The emergence of the VC LRO in a spin-gapped state is consistent with the emergent ferroelectricity in the absence of any spin-spiral magnetic LRO in experiments~\cite{Yasui13}. With further increasing magnetic field, the Zeeman interaction induces a finite magnetization most likely accompanied by the VC LRO and incommensurate spin spiral quasi-LRO, which also agrees with an anomaly in the dielectric constant and the specific heat in a recent experiment~\cite{exp}. However, it is still not clear how to understand the phenomenon at a quantitative level. The magnitude of the scalar chiral interaction, which is required for elucidating the experiments quantitatively, depends on how close to the D$_-$-VCD$_-$ boundary the system is located. This is still beyond the scope of this paper. The field-induced ring-exchange interaction might work more effectively between two neighboring spin chains, though we only considered it within a single chain for simplicity. It is also possible that the biquadratic Dzyaloshinskii-Moriya interaction induced by transverse optical phonon modes~\cite{onoda07} might assist the emergence of the vector-chiral LRO~\cite{furukawa08} with a smaller magnitude of the staggered scalar chiral interaction. A direct observation of the emergence or absence of the edge spin polarization at the edge in a particular boundary configuration will be useful to experimentally address the issue of whether the zero-field state is in the Haldane or even-parity dimer phase.

\begin{acknowledgements}

The authors thank S. Furukawa, F. Pollmann, M. Oshikawa, Y. Yasui, T. Kimura, I. Terasaki, T. Masuda, K. Tomiyasu, M. Hase, and R. Chitra for stimulating discussions. A part of calculations was performed by using the RIKEN Integrated Cluster of Clusters (RICC) facility. The work was partially supported by Grants-in-Aid for Scientific Research under Grant No. 24740253, 25800221 from Japan Society for the Promotion of Science.

\end{acknowledgements}

\newcommand{\PRL}[3]{Phys.\ Rev.\ Lett.\ {\bf #1}, #2 (#3)}
\newcommand{\PRLp}[3]{Phys.\ Rev.\ Lett.\ {\bf #1}, #2 (#3)}
\newcommand{\PRA}[3]{Phys.\ Rev.\ A {\bf #1}, #2 (#3)}
\newcommand{\PRAp}[3]{Phys.\ Rev.\ A {\bf #1}, #2 (#3)}
\newcommand{\PRB}[3]{Phys.\ Rev.\ B {\bf #1}, #2 (#3)}
\newcommand{\PRBp}[3]{Phys.\ Rev.\ B {\bf #1}, #2 (#3)}
\newcommand{\PRBR}[3]{Phys.\ Rev.\ B {\bf #1}, #2 (R) (#3)}
\newcommand{\PRBRp}[3]{Phys.\ Rev.\ B {\bf #1}, R#2 (#3)}
\newcommand{\arXiv}[1]{arXiv:#1}
\newcommand{\condmat}[1]{cond-mat/#1}
\newcommand{\JPSJ}[3]{J. Phys.\ Soc.\ Jpn.\ {\bf #1}, #2 (#3)}
\newcommand{\PTPS}[3]{Prog.\ Theor.\ Phys.\ Suppl.\ {\bf #1}, #2 (#3)}


\appendix

\section{Spatial averaging procedure for order parameters}
\label{average_process}
\begin{figure}[htb]
  \centering
  \resizebox{8.6cm}{!}{\includegraphics{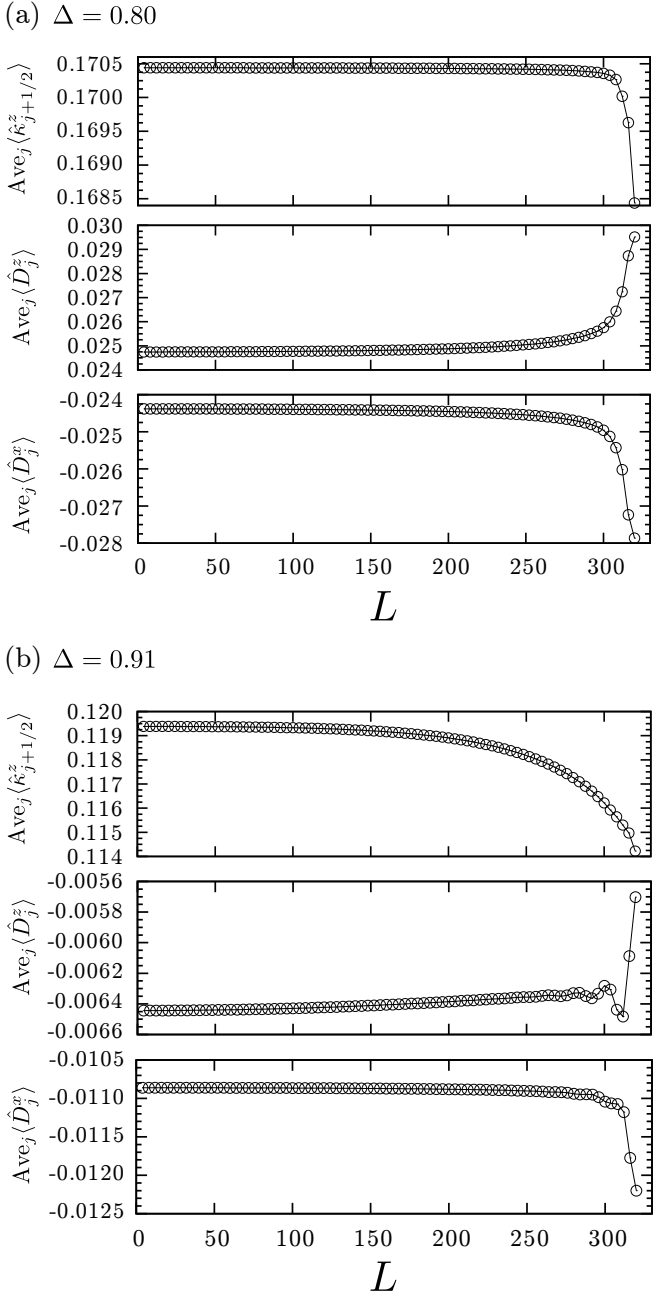}}
  \caption{ 
Spatially averaged order parameters for (a) $\Delta=0.8$ and (b) $\Delta=0.91$, where $J_1/J_2=-2.5$, $\delta=0.02$, $N=320$ and $m=500$. The parameter $L$ represents the number of sites over which the spatial average is taken.
}
\label{avg}
\end{figure}

When we compute order parameters, the results may show a spatial dependence, in particular, near the edges in the open boundary condition. This open boundary effect extends up to the longest length scale $\xi$ of the model, e.g., spin correlation length. This can be eliminated by taking the spatial averages of order parameters only over a certain number $L$ of spins in the middle of the spin chain, unless $\xi \ge (N-L)/2$. In Fig.~\ref{avg}, we show the $L$ dependence of order parameters for (a) $\Delta = 0.8$ and (b) $\Delta = 0.91$, where $J_1/J_2=-2.5$, $\delta=0.02$, and $N=320$. We take the maximum difference within the central region of the length $L$ as an error of the estimate. By taking $L=N/2$, we can reduce the boundary effects as much as possible while reserving a large enough number of sites for averaging.

\section{$N \rightarrow \infty$ extrapolation of the vector-chirality order parameter near the criticalities}
\label{fss_vc}
Here, we explain how we estimate the D$_-$-VCD$_-$ and D$_+$-VCD$_+$ critical points and interpolate the VC order parameter near the criticalities, as shown with broken curves in Fig.~\ref{fig:scaling} (a).

\begin{figure}[htb]
  \centering
  \resizebox{8.6cm}{!}{\includegraphics{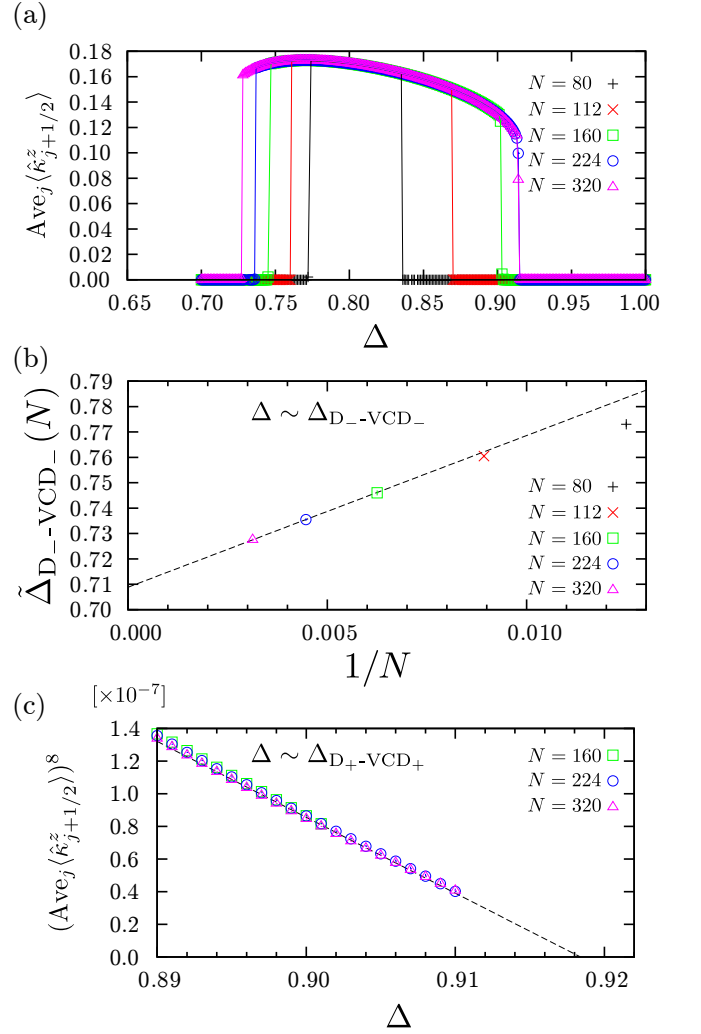}}
  \caption{ (Color online) 
(a) Vector-chirality order parameter ${\rm Ave}_j \langle \hat{\kappa}^{z}_{j+1/2} \rangle$ for $N$ up to 320 ($J_1/J_2=-2.5$ and  $\delta=0.02$). (b) The $N$ dependence of $\tilde{\Delta}_{ {\rm D}_- {\rm \mathchar`-VCD}_-}$. The broken line represents a linear extrapolation of the two largest-$N$ results. (c) $( {\rm Ave}_j \langle \hat{\kappa}^{z}_{j+1/2} \rangle )^8$ versus $\Delta$ near the $\Delta_{ {\rm D}_+ {\rm \mathchar`-VCD}_+}$ phase boundary. The broken line represents the least-square fit. All the results are obtained with $m=500$ and are unchanged within the symbol size by taking $m=400$.
}
\label{scaling_op_vc}
\end{figure}

Figure \ref{scaling_op_vc} (a) shows the VC order parameter versus $\Delta$ for $N$ up to 320 ($J_1/J_2=-2.5$ and $\delta=0.02$). It is clear that the region with ${\rm Ave}_j \langle \hat{\kappa}^{z}_{j+1/2} \rangle \neq 0$ becomes wider with increasing $N$. In particular, $\tilde{\Delta}_{ {\rm D}_- {\rm \mathchar`-VCD}_-}(N)$, defined as the valued of $\Delta$ at which ${\rm Ave}_j \langle \hat{\kappa}^{z}_{j+1/2} \rangle$ decays to zero with decreasing $\Delta$ around the D$_-$-VCD$_-$ phase boundary, shifts to a smaller $\Delta$. As shown in Fig.~\ref{scaling_op_vc} (b), $\tilde{\Delta}_{ {\rm D}_- {\rm \mathchar`-VCD}_-}(N)$ approaches the value 0.709 which defines the D$_{-}$-VCD$_-$ critical point $\Delta_{ {\rm D}_- {\rm \mathchar`-VCD}_-}$. Actually, the central charge in the conformal field theory at $\Delta_{ {\rm D}_- {\rm \mathchar`-VCD}_-}$ has been determined to be $1/2$ in Sec. \ref{sec:0criticality}, indicating the Ising criticality with the critical exponent $\beta=1/8$. Therefore, we interpolated $\Delta=\Delta_{ {\rm D}_- {\rm \mathchar`-VCD}_-}$ with ${\rm Ave}_j\langle \hat{\kappa}^{z}_{j+1/2} \rangle = 0$ and the closest point with ${\rm Ave}_j\langle \hat{\kappa}^{z}_{j+1/2} \rangle \neq 0$ by a functional form of $a(\Delta - \Delta_{ {\rm D}_- {\rm \mathchar`-VCD}_-})^{\beta}$ with $a$ being a nonuniversal constant.

Figure \ref{scaling_op_vc} (c) shows $( {\rm Ave}_j \langle \hat{\kappa}^{z}_{j+1/2} \rangle )^8$ as a function of $\Delta$ around $\Delta_{ {\rm D}_+ {\rm \mathchar`-VCD}_+}$. The broken line represents the least-square fit. This linear behavior of $( {\rm Ave}_j \langle \hat{\kappa}^{z}_{j+1/2} \rangle )^8 \propto (\Delta_{ {\rm D}_+ {\rm \mathchar`-VCD}_+} - \Delta)$ suggests that this transition belong to the Ising universality class. From the extrapolation to $N \rightarrow \infty$, we conclude the D$_+$-VCD$_+$ phase boundary is located at $\Delta_{ {\rm D}_+ {\rm \mathchar`-VCD}_+} \simeq 0.918$. This value of $\Delta_{ {\rm D}_+ {\rm \mathchar`-VCD}_+}$ reasonably agrees with $0.92$, which is obtained from the analysis of correlation functions shown in Fig.~\ref{fig:col_V=h=0} of Sec. \ref{sec:0mag:correlation}.

\end{document}